\documentclass[graybox, envcountchap]{svmult}

\usepackage{mathptmx}        
\usepackage{amsmath}
\usepackage{amssymb}
\usepackage{color}
\usepackage{helvet}          
\usepackage{courier}         
\usepackage{dirtree}

\usepackage{makeidx}        
\usepackage{graphicx} 
\usepackage{subfig}
\usepackage[driverfallback=dvipdfm]{hyperref}
\usepackage{multicol}        
\usepackage[bottom]{footmisc}

\hypersetup{colorlinks=true,urlcolor=blue}
\usepackage{wrapfig}
\usepackage[misc]{ifsym}

\makeindex             

\begin{document}

%

\title{Hitomi/XRISM micro-calorimeter}
\author{Kosuke Sato, Yuusuke Uchida, and Kumi Ishikawa on behalf of the Resolve team}
\institute{First Author (\Letter) \at Kosuke Sato, Saitama University, Japan, \email{ksksato@phy.saitama-u.ac.jp}
\and Second Author \at Yuusuke Uchida, Tokyo University of Science, Japan, \email{yuuchida@rs.tus.ac.jp}
\and Third Author \at Kumi Ishikawa, Tokyo Metropolitan University, Japan \email{kumi@tmu.ac.jp}}
%
%
\maketitle

\abstract{
We present an overview of the ASTRO-H (Hitomi) Soft X-Ray Spectrometer (SXS) and the X-Ray Imaging and Spectrometer Mission (XRISM) {\it Resolve} spectrometer.
In each, a 36-pixel X-ray micro-calorimeter array operated at 50 mK covers a $3 \times 3$ arc-minute field of view. The instruments are designed to achieve an energy resolution of better than 7 eV over the 0.3 -- 12 keV energy range and operate for more than 3 years in orbit. Actually, the SXS achieved the energy resolution of $\sim$5 eV in orbit, but it was lost after only a month of operation due to the loss of spacecraft attitude control. For the recovery mission, XRISM will be equipped with the {\it Resolve} spectrometer which has mostly the same design as SXS and is expected to have the same in-flight performance.
}

\clearpage

\section*{Scope}

This chapter describes the micro-calorimeter systems, SXS and {\it Resolve} onboard ASTRO-H (Hitomi) and X-Ray Imaging and Spectroscopy Mission (XRISM), respectively. The design of these instruments is basically the same with a few changes. This chapter is also intended to help{\it Resolve} users understand the basic principles and configurations of this instrument for data analysis. This also serves as a gateway to a collection of references for more technical details on specific topics. The structure of this chapter is as follows. An introduction to and motivation for the micro-calorimeter are described in Section \ref{sec:intro}.
We explain the configuration of the instruments such as the detector, onboard event processing, and cooling systems in Section \ref{sec:instrument}. The data processing, screening, and calibrations on the ground tests are described in Section \ref{sec:process}. In Section \ref{sec:hitomi-results}, we report the Hitomi/SXS performance in orbit and the observations for the Perseus cluster. Finally, we describe the performance of  XRISM/{\it Resolve} achieved in the ground test in Section \ref{sec:resolve}. 

\section{Introduction}
\label{sec:intro}

X-ray micro-calorimeters detect the energy of incident X-rays by converting the energy of each photon into heat in an absorber with small heat capacity attached to a highly sensitive thermometer \cite{Moseley1984, Kelley1999, Kelley2007, Kilbourne2018b}. In principle, a high resolution of $E/\Delta E >$1000 can be obtained, and the high resolution can be achieved in soft X-ray to hard X-ray ranges. X-ray micro-calorimeters are non-dispersive detectors. Compared to dispersive detectors such as grating spectrometers, which provide high energy resolution below about 1 keV, their energy resolution does not deteriorate over a whole energy range and their performance does not degrade for diffuse objects such as clusters of galaxies.  Thus, they have the potential to investigate many phenomena in high-energy astrophysics.

\begin{figure}[b]
\centering
\includegraphics[scale=0.4]{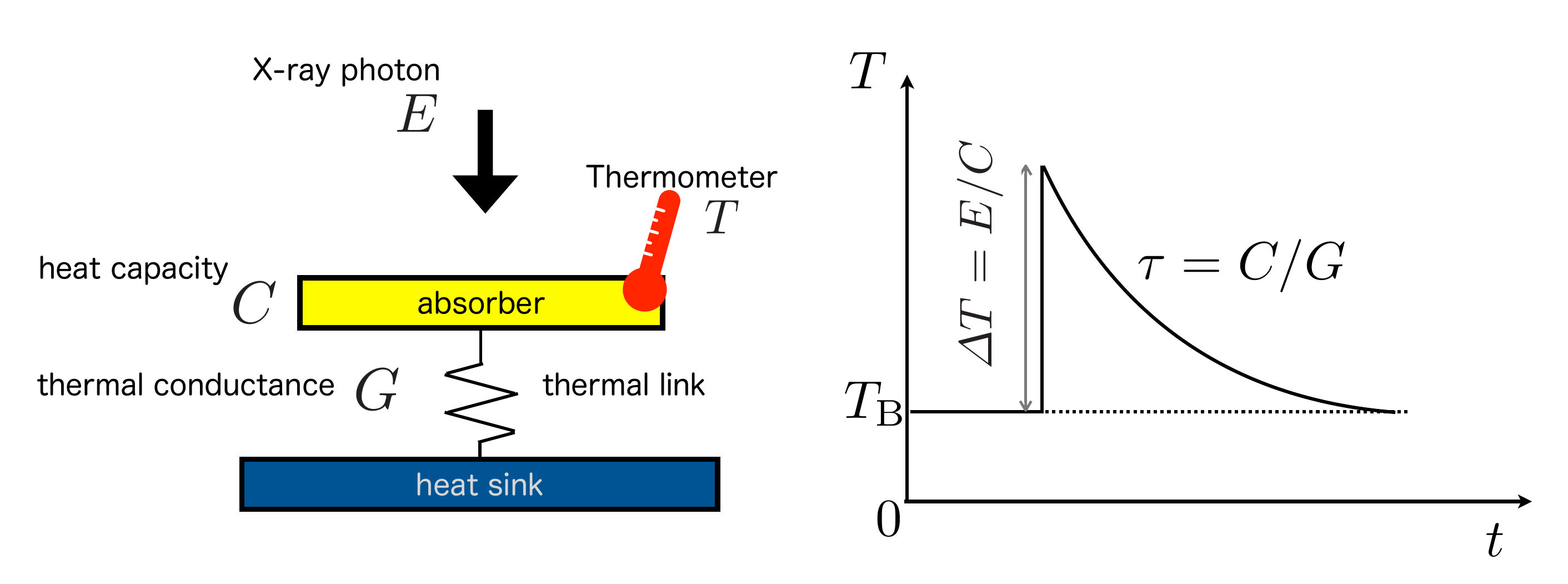}
\caption{Schematics of micro-calorimeter.}
\label{fig:calorimeter}
\end{figure}

X-ray micro-calorimeters are thermal sensors that detect the energy of incident X-ray photons as a small temperature rise in the sensor. An X-ray micro-calorimeter consists of an absorber, thermometer, thermal link, and heat sink as shown in Figure \ref{fig:calorimeter}. An absorbed X-ray photon transfers energy to a photo-electron that then imparts energy to electrons and phonons in a process called thermalization.  Ideally, a quasi-equilibrium state is achieved, and the resulting distribution of energy can be characterized by a temperature.  This small temperature change $\Delta T$ can be expressed approximately by $\Delta T = E / C$, where the incident energy is $E$ and the heat capacity of the sensor is $C$. To avoid position-dependent effects, equilibration in the absorber must be faster than the equilibration time of the thermometer with the absorber. The heat generated in the absorber escapes through the thermal link to the heat sink. The temperature returns to the steady state by the time constant of $\tau \sim C/G$, where $G$ is thermal conductance. The energy resolution of the micro-calorimeter is expressed \cite{Irwin1995} as
\begin{equation}
\Delta E \propto \sqrt{k_\mathrm{B}T^2 C / \alpha}, \nonumber
\end{equation}
where $k_\mathrm{B}$ is the Boltzmann constant, $T$ is a temperature of absorber, and $\alpha = \mathrm{d}\ln R / \mathrm{d}\ln T$ is the logarithmic sensitivity of the thermometer \cite{Kilbourne1999}.
The equation indicates that $\Delta E$ is constant over the energy, in contrast to the grating spectrometer with constant $\Delta \lambda$ over wavelengths.

\begin{figure}[b]
\centering
\includegraphics[scale=.3]{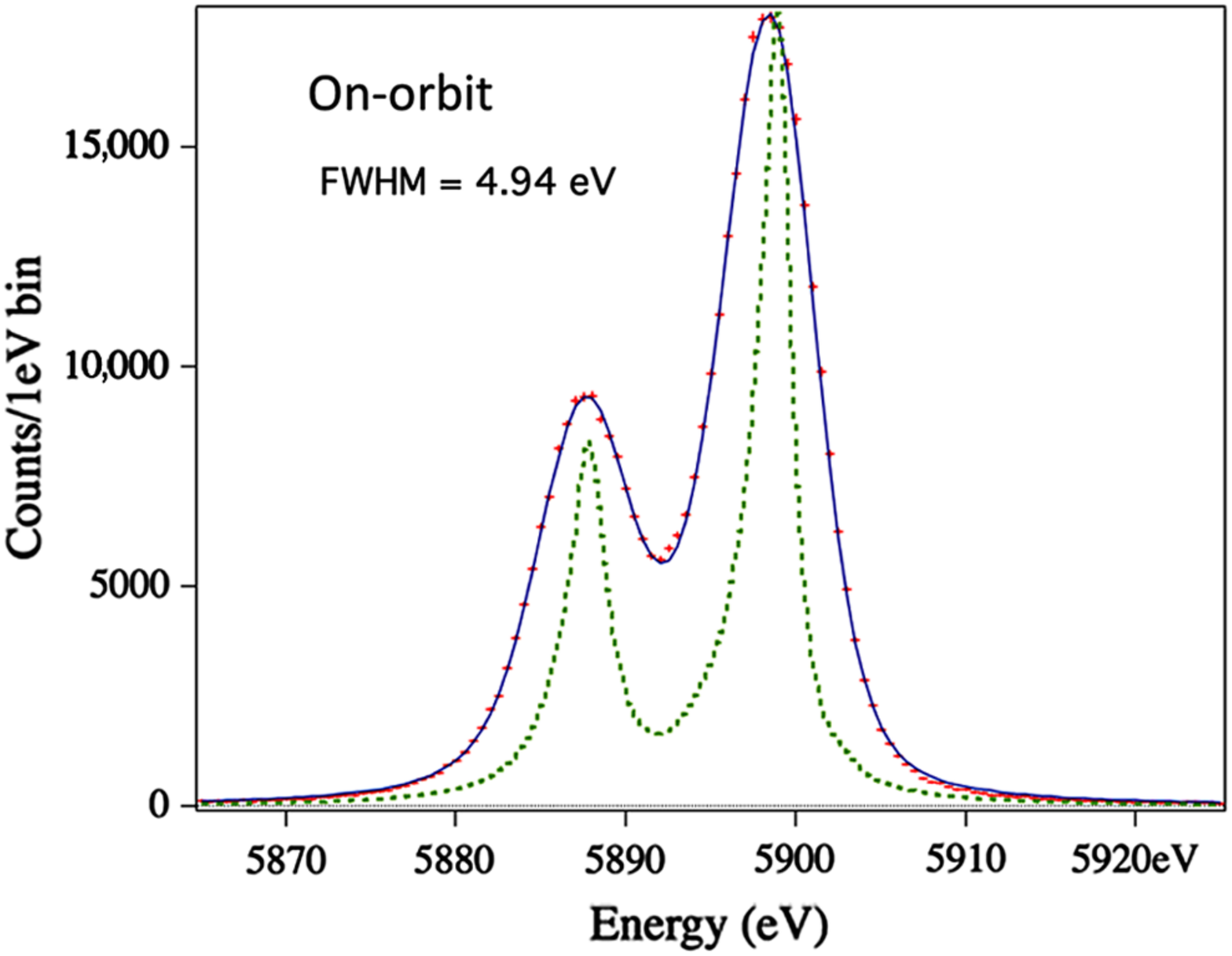}
\caption{The SXS performance in orbit \cite{Porter2018}. The spectra of Mn K$\alpha$ at 5.9 keV for all the pixels were added together. [Reproduced with permission from Porter, F. S., Boyce, K. R., Chiao, M. P., et al., Journal of Astronomical Telescopes, Instruments, and Systems, 4, 011218 (2018). Copyright 2018 Author(s), licensed under a Creative Commons Attribution 4.0 License.]}
\label{fig:resolution}       
\end{figure}

\begin{table}[t]
\caption{Requirements for Hitomi/SXS and XRISM/{\it Resolve} as shown in \cite{Ishisaki2022}. [Reproduced with permission from Ishisaki, Y., et al., Proc. SPIE Int. Soc. Opt. Eng., 12181, 121811S (2022)]}
\label{tab:resolve-req}
\includegraphics[width=\linewidth]{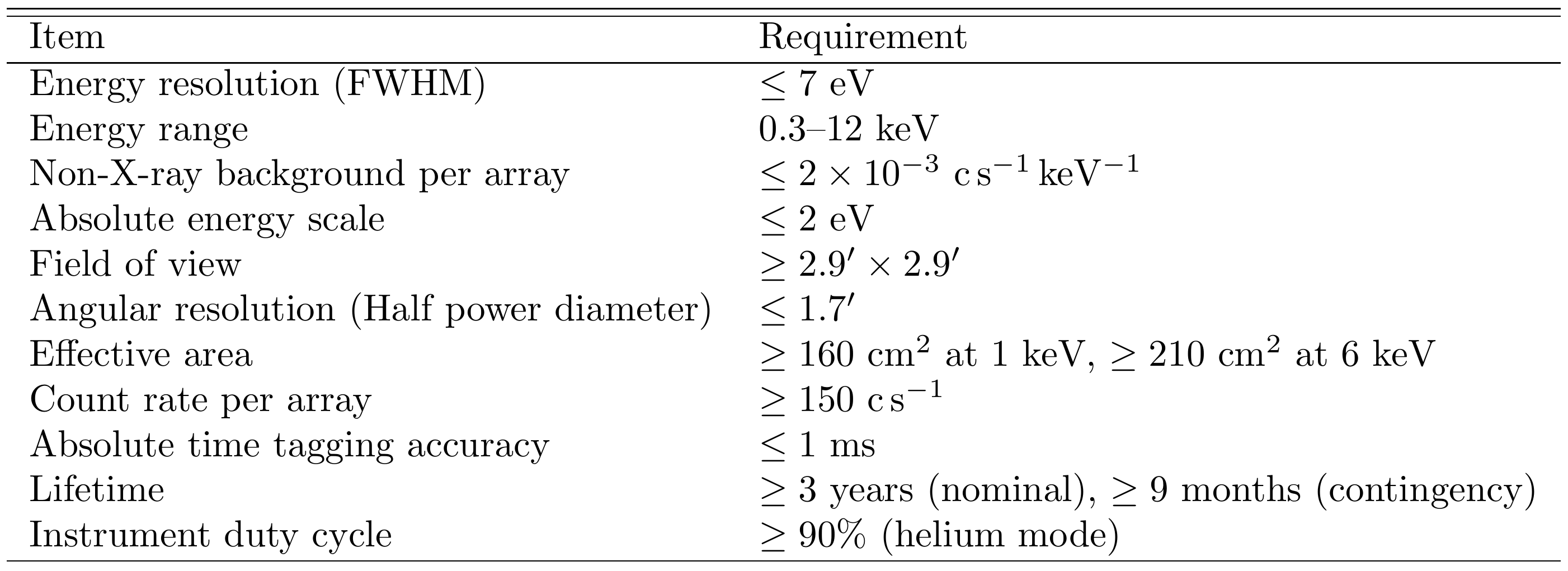}
\end{table}

\begin{figure}[b]
\centering
\includegraphics[scale=.4]{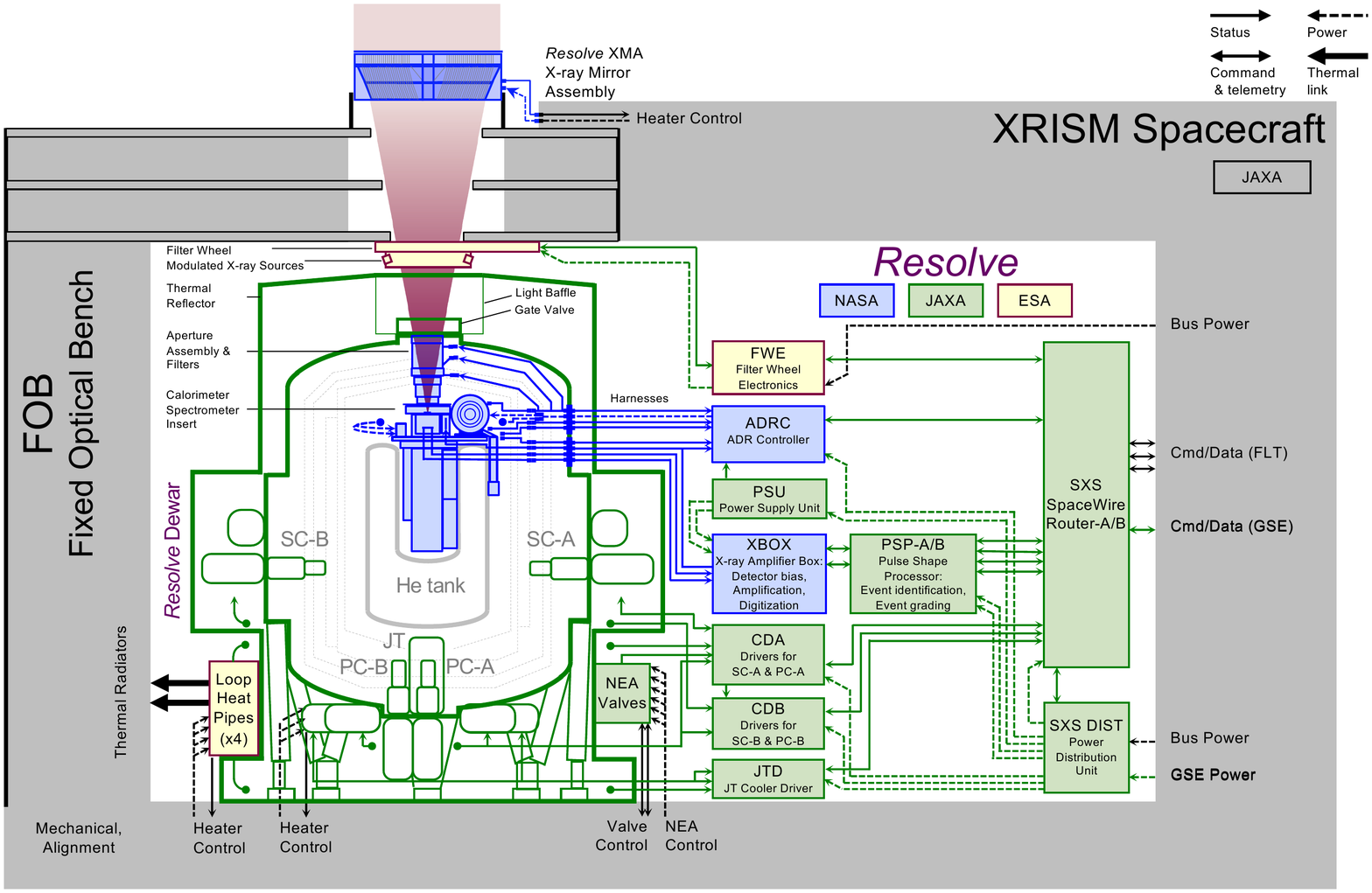}
\caption{Block diagram of the {\it Resolve} instrument \cite{Ishisaki2022}. Details are found in the text. [Reproduced with permission from Ishisaki, Y., et al., Proc. SPIE Int. Soc. Opt. Eng., 12181, 121811S (2022)]}
\label{fig:block-diagram}       
\end{figure}

Historically, the idea of the X-ray micro-calorimeter originated as a search for a material with semiconductor band-gap smaller than Si to enable higher spectral resolution in semiconductor devices. This led to the identification of candidate materials that are commonly used in infrared bolometers, and this eventually lead to the suggestion \cite{Moseley1984} of using a bolometer-type detector to thermally detect the energies of individual X-ray photons. This concept was initially proposed for the AXAF mission (now named Chandra) by NASA (National Aeronautics and Space Administration), but restructuring of that program led to the adoption of a micro-calorimeter spectrometer, called the X-ray Spectrometer (XRS), on the ASTRO-E mission \cite{Kelley1999}, which was the launch of ASTRO-E in 2000 by ISAS (Institute of Space and Astronautical Science) and NASA, but it was not put into orbit due to a rocket malfunction. On the other hand, this XRS-type sensor has been used in many ground-based (e.g., \cite{Brown2006}) and rocket experiments with success such as the X-Ray Quantum Calorimeter (XQC) \cite{McCammon2002}. The recovery mission of ASTRO-E, Suzaku \cite{Mitsuda2007}, was successfully launched, and the in-orbit performance (e.g., the energy resolution of the calibration pixel) basically achieved in the early phase after launch \cite{Kelley2007}. However, the He cryogen was lost and the operations of XRS were terminated before observing celestial objects. 

\begin{figure}[b]
\centering
\includegraphics[scale=.25]{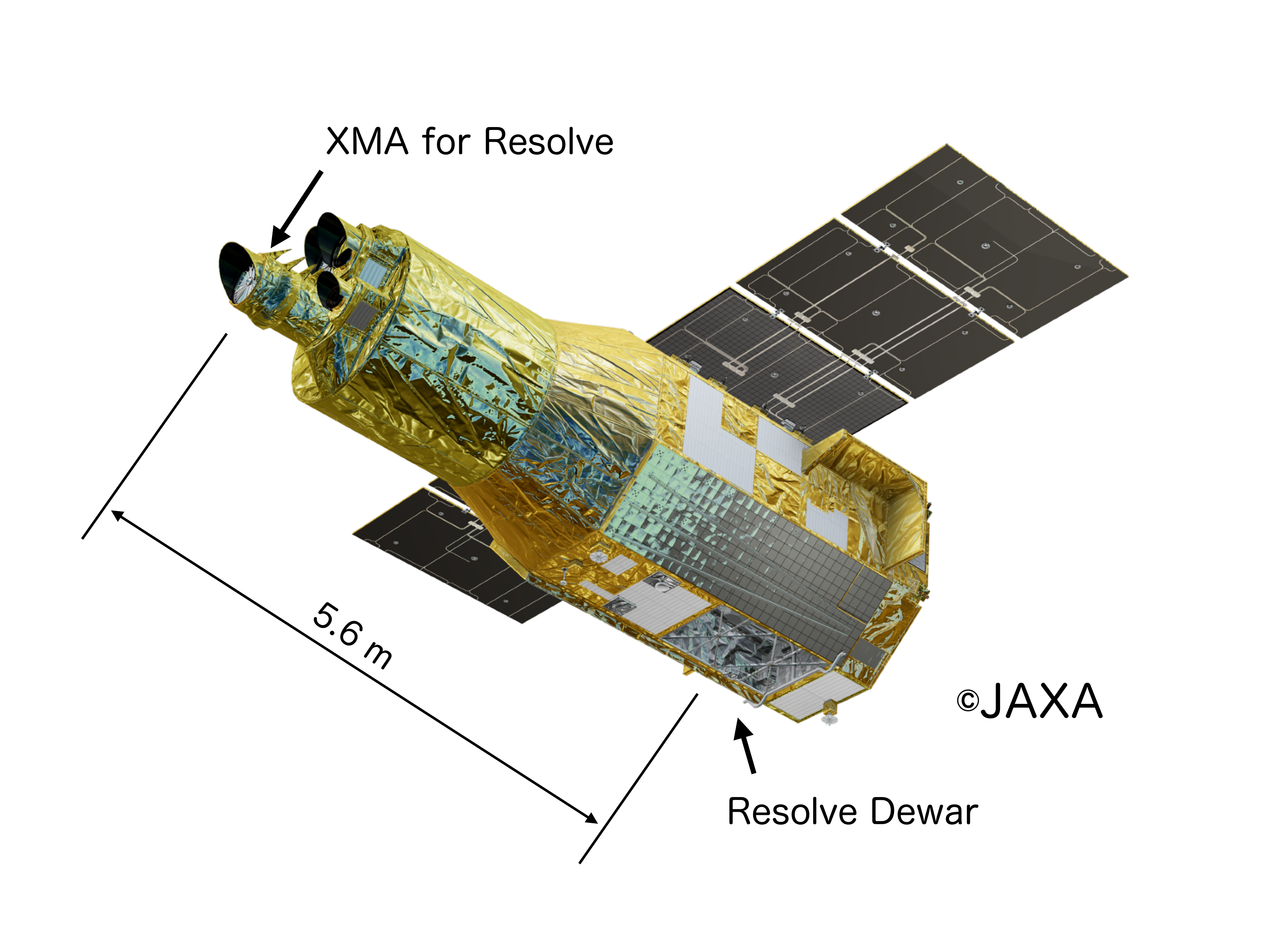}
\caption{Schematic view of the XRISM spacecraft.}
\label{fig:xrism}       
\end{figure}

The ASTRO-H (Hitomi) was launched on February 17, 2016, as the 6th Japanese X-ray astronomy satellite and developed under the international collaboration of JAXA (Japan Aerospace Exploration Agency), NASA, ESA (European Space Agency), and CSA (Canadian Space Agency) \cite{Takahashi2018}. The Hitomi SXS is a system that combines an X-ray micro-calorimeter spectrometer with a Soft X-ray Telescope (SXT) to cover a $3' \times 3'$ field of view (FOV) with an angular resolution of $1.7'$ (half power diameter) \cite{Kelley2016, Okajima2016} as shown in Table \ref{tab:resolve-req}. The micro-calorimeter detector is an array of 6 $\times$ 6 pixels, arranged at an 832 ${\rm \mu m}$ pitch \cite{Kilbourne2018b}. It was operated at 50 mK under a stable environment inside a dewar with a multi-stage cooling system as described in Section \ref{subsec:cooling} in detail. The SXS was designed to achieve an energy resolution of better than 7 eV (FWHM) at 6 keV and has actually achieved 5 eV (FWHM) at 6 keV in orbit as shown in Figure  \ref{fig:resolution}. However, the Hitomi had an accident with the attitude control system in March 2016, about a month after the launch. Thus, only the X-ray spectra from a few celestial objects in relatively higher energy bands could be observed because the gate valve on the SXS dewar, which consists of a Be window and its support structure had not been opened.

The design of {\it Resolve} onboard XRISM is basically the same as the Hitomi SXS \footnote{\url{https://xrism.isas.jaxa.jp/research/analysis/manuals/xrqr_v1.pdf}} with a few changes. See Section 3 in \cite{Ishisaki2022} for the details. Figure \ref{fig:block-diagram} shows a block diagram of the {\it Resolve}. 
The dewar for the {\it Resolve} is located on the base panel, while the X-ray Mirror Assembly (XMA) is mounted on top of the optical bench, as shown in Figure \ref{fig:xrism} \cite{Ishisaki2022}.

\section{Instruments of Hitomi/SXS and XRISM/{\it Resolve}}
\label{sec:instrument}

This section describes each component in the {\it Resolve} instrument. The detector system, micro-calorimeter, and anti-coincidence detector are described in Section \ref{subsec:detector}. We explain the signal processing chain to read X-ray events with the analog and digital electronics box in Section \ref{subsec:event_processing}, and the characteristics, such as the read-out system and energy spectrum, of the micro-calorimeter in Section \ref{subsec:characteristics}. The cooling chain to cool the detector to 50 mK with He cryogen, mechanical coolers, and adiabatic demagnetization refrigerators (ADR) in the He dewar is shown in Section \ref{subsec:cooling}. We describe the optical chain through the X-ray path in the He dewar including filter wheel (FW), aperture assembly (ApA), blocking filters, and gate valve (GV) in Section \ref{subsec:xray-path}. Section \ref{subsec:cal-sc} describes the onboard calibration sources, to correct temporal energy gain variation later on the ground.

\subsection{Detector system}
\label{subsec:detector}

\begin{figure}[b]
\centering
\includegraphics[width=\linewidth]{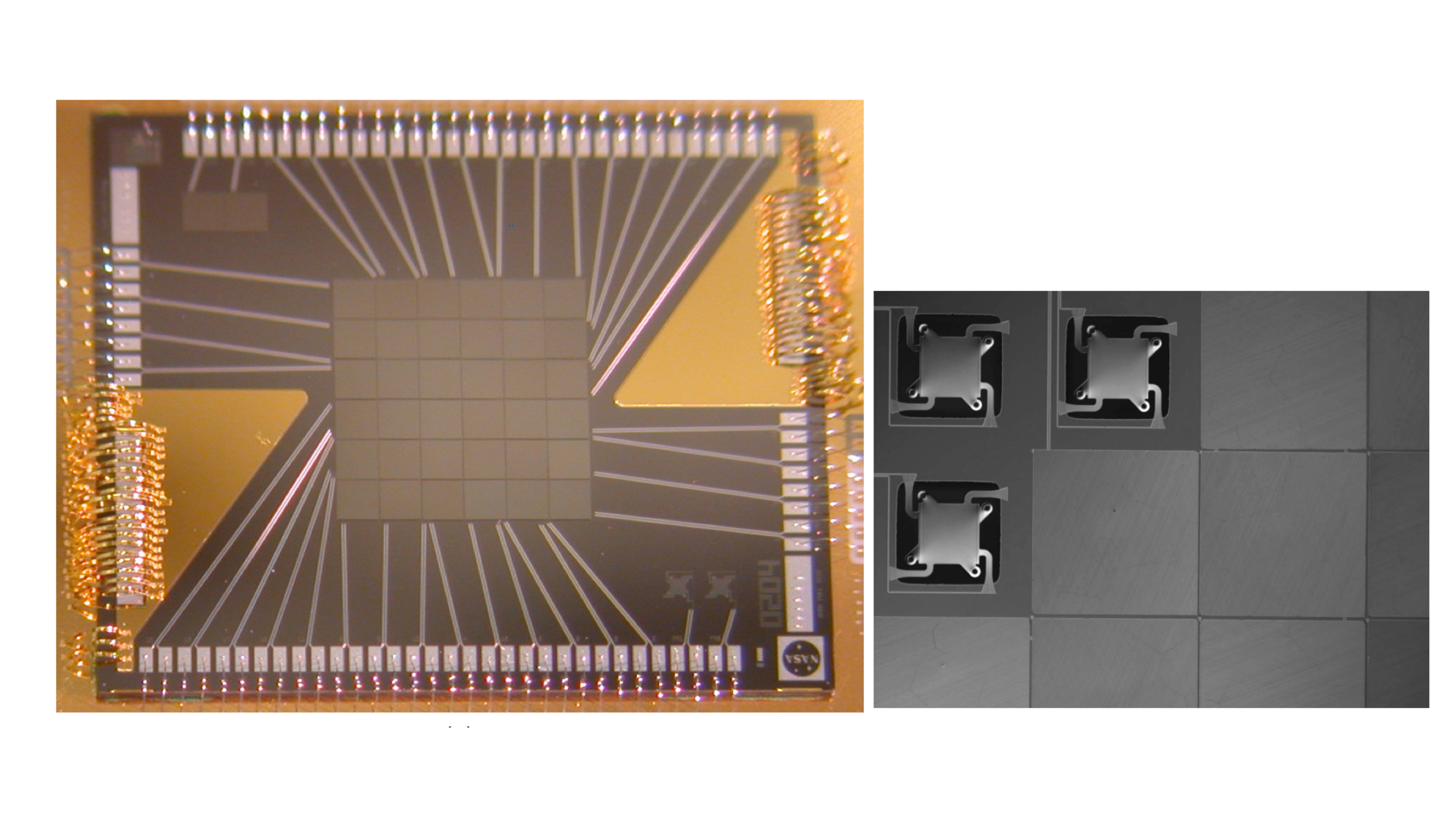}
\caption{(Left): SXS spare array. An array of 6 x 6 pixels comprises a field of view, and the upper left pixel outside of the FOV is a calibration pixel. (Right): Portion of a flight-candidate array before completing the attachment of the X-ray absorbers to the pixels.  The thermally isolated thermistors are visible on the pixels without absorbers \cite{Kilbourne2018b}. [Reproduced with permission from Kilbourne et al., J. Astron. Telesc., Instrum., Syst. 4(1), 011214 (2018). Copyright 2018 Author(s), licensed under a Creative Commons Attribution 4.0 License.]}
\label{fig:sxs-pix-abs}       
\end{figure}

The SXS and {\it Resolve} detector assembly is basically the same: a 36-pixel X-ray micro-calorimeter array whose pixels consist of ion-implanted Si thermistors with HgTe absorbers at an 832 ${\rm \mu}$m pitch as shown in the left panel in Figure \ref{fig:sxs-pix-abs} \cite{Chiao2018, Kilbourne2018b}. One upper left corner of the array, outside the FOV of the X-ray mirror in Figure \ref{fig:sxs-pix-abs}, is a calibration pixel.  This pixel is illuminated by a collimated $^{55}$Fe source which continuously provides monitoring of the gain scale and the line spread function (LSF). The array covers a field of view of $3' \times 3'$ and an energy range of 0.3 -- 12 keV\@. The requirements are shown in Table \ref{tab:resolve-req}, and the energy resolution in orbit actually achieved 5 eV at 6 keV\@ by the SXS onboard Hitomi, as mentioned above. 

As high-sensitivity thermometers, semiconductor thermistors that operate in the variable-range-hopping conduction regime have been often used in astrophysics \cite{McCammon2005, Kelley2007, Kilbourne2018b}. Particularly, the resistance, $R$, of an ion-implanted Si thermistor at low temperatures, $T$, in the variable-range-hopping regime with the Coulomb gap can be expressed as $R(T) = R_0 \exp{[(T_0/T)^{0.5}]}$, where $R_0$ and $T_0$ are constants. Ion-implanted Si often deviates from this equation, but performance can be predicted in the small-signal regime from the logarithmic sensitivity $\alpha$ at the operating temperature. In the case of the SXS detector, $\alpha$ is about $-6.3$ at the detector temperature under bias. The time constant of the falling time (i.e., the time to return to a steady state) after an X-ray absorbed photon is 3.5 msec \cite{Kilbourne2018b}. The absorber material should have a small heat capacity and a large stopping power of X-rays to increase the temperature change during energy injection. A semimetal, HgTe, has been used as an absorber as a material that satisfies those conditions since the XRS. The HgTe absorbers are attached with epoxy on each pixel as shown in the right panel in Figure \ref{fig:sxs-pix-abs}. Both the designed quantum efficiency at 6 keV and a filling factor of the SXS were greater than 95\%\@. This type of micro-calorimeter typically has high electrical impedance. Due to the coupling capacitance of the long harnesses, the high impedance makes the system extremely sensitive to microphonics. Thus, a junction field-effect transistor (JFET) source follower circuit has been applied to convert the high impedance of the detector to the low output impedance of the JFET  \cite{Kelley2007}. 

The SXS and the {\it Resolve} also have the same anti-coincidence (anti-co) detector as the XRS to exclude cosmic-ray events and to monitor the particle environment \cite{Kelley2007, Kilbourne2018b}. The cosmic rays hit both the detector pixel and anti-co and hence the background events due to cosmic rays can be removed by anti-coincidence. The silicon ionization detector is used for the anti-co detector, which is located behind the micro-calorimeter detector array. The anti-co detector consists of a 1 cm$^2~\times 0.5$ mm high purity silicon configured as a p-i-n diode, and the anti-co detector covers a larger area than the micro-calorimeter detector.
In the case of the SXS, the pulse falling time and the dead time were 0.15 msec and $<1$ msec, respectively, and the signal was much faster than the micro-calorimeter signal (3.5 msec).  

\subsection{Event processing system}
\label{subsec:event_processing}

Event processing, using a filter to maximize the signal-to-noise ratio, is essential to achieve the high spectral performance of the micro-calorimeters. The optimal filtering \cite{Boyce1999} is thus adopted. In the optimal filtering, a template is created for each pixel from the responsivity calculated by an average pulse and the noise spectrum. The pulse height is calculated by cross-correlation between the optimal filter template and the waveform of each X-ray event. 
If another X-ray event (pulse) contaminates the waveform, the pulse height would not be calculated correctly. Therefore, the contamination of another event should be detected accurately.
The derivatives of time series data are examined to detect the contamination of another pulse because they are more sensitive than the time series data. It is necessary to discriminate grades according to the time interval to the preceding and following pulses and to change the pulse processing for each grade. The pulse shape processor (PSP) of the SXS and the {\it Resolve} instruments is designed to implement these functions. 

\begin{figure}[b]
\centering
\includegraphics[width=0.95\linewidth]{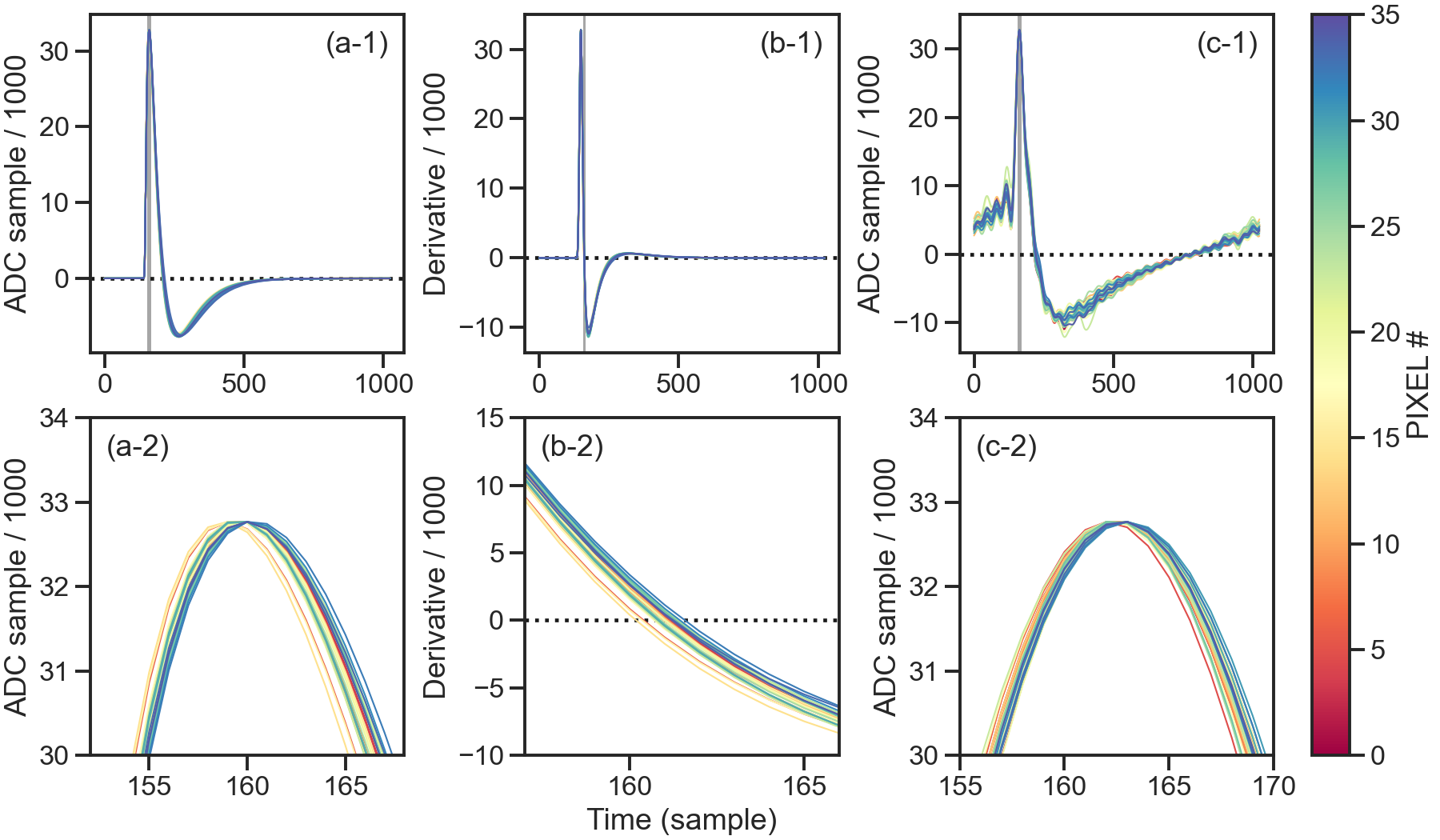}
\caption{%
Pulse profiles (a-1), profiles of its time derivatives (b-1), and the templates of high-resolution grade (Hp) (c-1) for each pixel in \cite{Omama2022}.
Figures (a-2), (b-2), and (c-2) show figures zoomed in around the peaking time (sample) of the pulse profiles for (a-1), (b-1), and (c-1), respectively.
There are time offsets among pixels, which are compensated by the template for each pixel. The pulse overshoots the baseline as shown in the upper panels due to the AC coupling in the XBOX. [Reproduced with permission from Omama, T., Tsujimoto, M., Sawada, M., et al., Proc. SPIE Int. Soc. Opt. Eng., 12181, 121861 (2022)]
}
\label{fig:adcsample}
\end{figure}

The signals of the micro-calorimeter pixels and the anti-co detector are amplified and digitized by the analog signal processing unit called the X-ray amplifier BOX (XBOX). The digitized data sampled with a sampling rate of 12.5 kHz by the analog-to-digital converter (ADC) in the XBOX, hereafter called \texttt{adc\_sample}, is transferred to the PSP for event detection and pulse height calculation. The PSP applies a boxcar filter to the \texttt{adc\_sample} and calculates the \texttt{derivative}. Figure \ref{fig:adcsample} (a) and (b) show the \texttt{adc\_sample} of the pulse shape and the \texttt{derivative} \cite{Omama2022}. Both the \texttt{adc\_sample} and the \texttt{derivative} are stored in a buffer memory named waveform ring buffer (WFRB) in the PSP \cite{Ishisaki2018}. 

\begin{figure}[bt]
\centering
\includegraphics[width=0.95\linewidth]{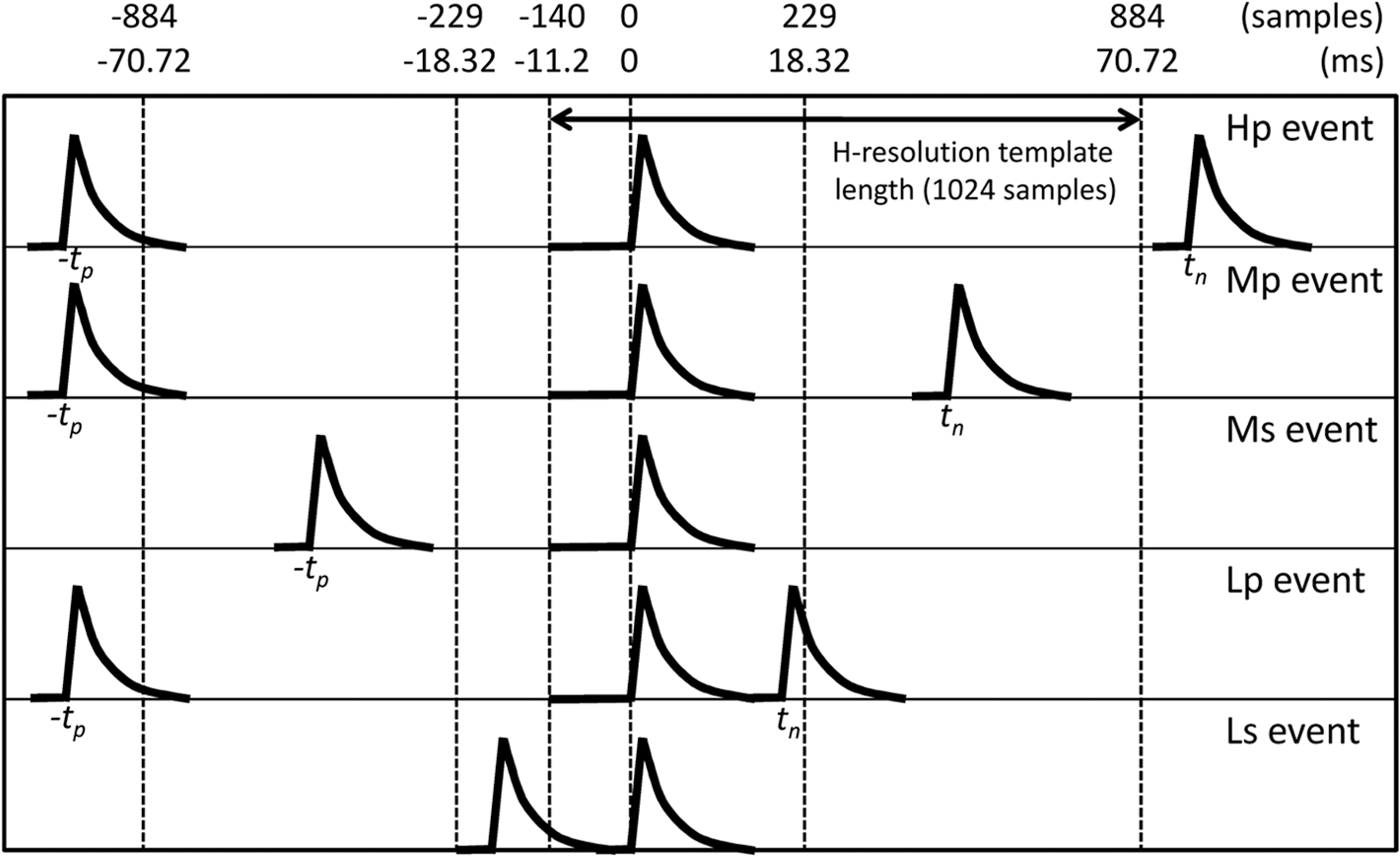}
\caption{%
Schematic view of grades for pulse arriving at $t = 0$ \cite{Ishisaki2018}. [Reproduced with permission from Ishisaki, Y., et al., Journal of Astronomical Telescopes, Instruments, and Systems, 4(1), 011217 (2018). Copyright 2018 Author(s), licensed under a Creative Commons Attribution 4.0 License.]
}
\label{fig:grade}
\end{figure}

The PSP triggers X-ray events, assigns an event grade, and performs optimal filtering to calculate pulse heights. An X-ray event is triggered when the \texttt{derivative} exceeds the event threshold determined for each pixel. Once triggered, the PSP searches for the following pulse by subtracting average \texttt{derivative} scaled to the primary pulse, and by checking if the residual exceeds the event threshold again. This process is needed because another pulse, if exists, contaminates the preceding pulse signal to introduce an error in the pulse height calculation. If there is no other pulse within $\pm 884$ samples (70.72 ms) with respect to the triggered time (see the top panel of Fig.\ \ref{fig:grade}), the triggered event is classified as high-resolution primary (Hp) grade. Regarding Hp grade events, the PSP adopts the record of 1024 (81.92 ms) including 140 preceding samples (see Fig.\ \ref{fig:adcsample}, a-1), and calculates the cross-correlation between this pulse record and the optimal filter template of the same length (Fig.\ \ref{fig:adcsample}, c-1). 
The maximum of the cross-correlation gives an estimate of the pulse height. The PSP searches for the maximum cross-correlation by shifting the pulse record. The final pulse height amplitude (PHA) and the photon arrival time are obtained at a sub-integer sample (1/16, which corresponds to 5 $\mu$sec resolution) of the sampling frequency, by using the cross-correlation values at $-1$, 0, and $+1$ shifts with respect to the shift that gives the maximum, and assuming the cross-correlation is locally represented by the second order polynomial of time near the maximum.
See \cite{Boyce1999} and \cite{Ishisaki2018} for further details on optimal filter processing.

If there is no preceding pulse from the $-884$ sample, and if there is a second pulse within the following 884 samples, but not within the following 229 samples (18.32 ms, see the second panel of Fig.\ \ref{fig:grade}), the triggered event is classified as medium-resolution primary (Mp) grade. 
As shown in Fig. \ref{fig:grade}, the time intervals to the preceding pulse and to the following pulse determine the grade.
Regarding the event of Mp or medium-resolution secondary (Ms) grade, the PSP adopts the record of 256 samples (20.48 ms) and determines its pulse height using the short template.

In a similar way, using the interval of $\pm 229$ samples (18.32 ms), the triggered event is classified as low-resolution primary (Lp) grade or low-resolution secondary (Ls) grade. The PSP does not apply the optimal filter to the events of Lp or Ls grades but calculates the pulse height by subtracting the baseline value from the maximum value of the \texttt{adc\_sample}.

\begin{table}[b]
\centering
\caption{%
The event grade definition \cite{Ishisaki2018}. $t_\mathrm{n}$ is the interval to the next pulse, while $t_\mathrm{p}$ is the interval to the previous pulse. [Reproduced with permission from Ishisaki, Y., et al., Journal of Astronomical Telescopes, Instruments, and Systems, 4(1), 011217 (2018). Copyright 2018 Author(s), licensed under a Creative Commons Attribution 4.0 License.]}
\begin{tabular}{lccc}
\hline\noalign{\smallskip}
& \ \ \ $t_\mathrm{p} \leq 18.32$\ \ \  & \ \ \ $18.32< t_\mathrm{p} \leq70.72$\ \ \  & \ \ \ $70.72 < t_\mathrm{p}$\ \ \  \\
\noalign{\smallskip}\hline\noalign{\smallskip}
$t_\mathrm{n} \leq 18.32$\ \ \  & Ls & Ls & Lp \\
$18.32 < t_\mathrm{n} \leq 70.72$\ \ \  & Ls & Ms & Mp \\
$70.72 < t_\mathrm{n}$\ \ \  & Ls & Ms & Hp \\
\noalign{\smallskip}\hline\noalign{\smallskip}
\end{tabular}
\label{tab:grade}
\end{table}

Table \ref{tab:grade} summarizes the relationship between the event grade and the time interval of the two pulses. See \cite{Ishisaki2018} for further details on the PSP implementation. The energy resolution of the Mp events is close to that of the Hp events, and hence, Hp and Mp events will be used in the standard data analysis.
 
The PSP has the capability to generate the templates onboard, as well as to upload pre-calculated templates on the ground. Templates that are used with the {\it Resolve} PSP in the spacecraft-level test, as well as the launch time, are those generated on the ground using the instrument-level test data (version of 2021 Dec 13), where the flight dewar and the flight electronics were used. 
High-frequency weight was removed for the flight templates with a cut-off frequency of 366 Hz because otherwise, the instrument could be vulnerable to increases in high-frequency noise that might occur after the optimal filter was calculated (for details, see Section 5.5 in \cite{Ishisaki2022}).

\subsection{Characteristics of the micro-calorimeters}
\label{subsec:characteristics}

In this section, we describe the evaluation of an absolute energy scale and energy resolution of incident X-ray photons to characterize the micro-calorimeter array. We also explain the effects of noise in the readout system and high-count observations, which affect the evaluation of these energy scales and energy resolutions.

The energy scale, i.e., the relation between the measured pulse height and the incident photon energy, is a key parameter to calculate accurate absolute measurements of photon energy. The SXS and the {\it Resolve} employ very sensitive thermometers and the energy scale is affected by these instruments' thermal environments. Different detector temperatures and radiative loads lead to changes in the energy scale. Therefore, it is necessary to remove temporal variations in the gain.
The SXS and the {\it Resolve} detector energy scale consists of two components: 1. all the pixels have a common gain that varies together with the detector's effective temperature which depends on the calorimeter thermal sink (CTS) temperature, 2. a differential gain among the pixels due to large temperature excursions or
changes in radiative loading \cite{Eckart2018, Leutenegger2018, Porter2018}. Details of the gain correction are described in Section \ref{ssec:gain}.

\begin{figure}[b]
    \centering
    \includegraphics[width=0.6\linewidth]{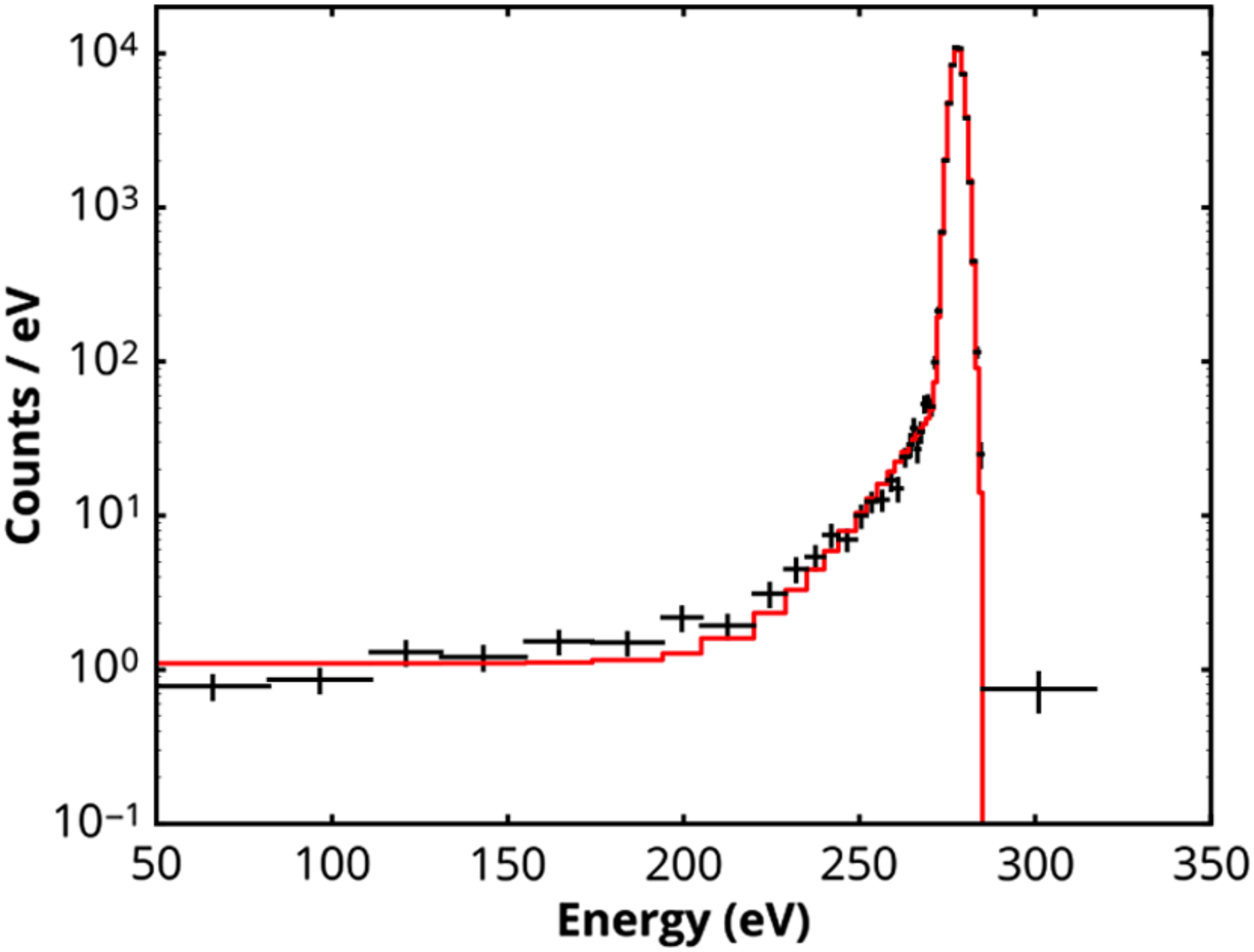}
    \caption{%
      The line-spread function (LSF) for a single SXS pixel in response to monochromatic X-rays at 282 eV \cite{Eckart2018}.
      The black points show data and the red curve shows the LSF model, illustrating three components: the Gaussian core which dominates the line shape,
      an exponential tail to low energies, and the electron loss continuum. [Reproduced with permission from Eckart, M. E., Adams, J. S., Boyce, K. R., et al., Journal of Astronomical Telescopes, Instruments, and Systems, 4, 021406 (2018). Copyright 2018 Author(s), licensed under a Creative Commons Attribution 4.0 License.]
}%
\label{fig:lsf}
\end{figure}

The high energy resolution is a fundamental science requirement of the SXS and the {\it Resolve} and is reflected in the LSF. The LSF for each pixel across the science bandpass was characterized by nearly monochromatic X-rays generated by several monochromators. An example of the LSF is shown in Figure \ref{fig:lsf}\@. The LSF is described with two components; a Gaussian core and an extended LSF which  includes an exponential tail with e-folding of $\sim$12 eV, an electron loss continuum, and escape peaks \cite{Eckart2018, Leutenegger2018}. The LSF core is well represented by a Gaussian down to at least three orders of magnitude. The extended LSF is made up of a small fraction of events redistributed to lower energies through several energy-loss mechanisms. These mechanisms include long-lived surface state excitation which gives rise to an exponential tail with an e-folding of about 12 eV. Scattering of photoelectrons from the absorber results in a so-called electron-loss continuum, which allows fluorescent X-rays to escape from the absorber instead of thermalizing, producing an escape peak. Because the Gaussian core is dependent on the detector and system noise, measurements on the ground can serve as a guide but must be re-measured in orbit.

Thermal and electrical crosstalk cause degradation of the energy resolution.
When an X-ray is absorbed in the frame of the micro-calorimeter chip, the frame event pulses are too small to trigger, but they add to the noise of each pixel. In addition, when an X-ray photon is absorbed in one pixel and that heat flows to the frame, a similar perturbation in the frame temperature occurs. This thermal crosstalk adds to the noise. In practice, this noise would have affected only observations of the bright celestial sources.  In orbit, cosmic rays are also sources of thermal crosstalk due to frame events, where minimum ionizing particles deposit energy into the silicon frame.
This effect leads to a slight degradation in energy resolution, with the magnitude of the effect dependent on the cutoff rigidity, which is a proxy for the rate of cosmic rays \cite{Leutenegger2018}. 

In high count rate situations, electrical crosstalk has a higher impact on the energy resolution and is considered to be a major cause of the degradation. In-ground tests, the degradation in energy resolution was confirmed to be a few eV \cite{Mizumoto2022}. 
The electrical cross-talk occurs between adjacent channels in the high-impedance part of the circuit prior to the JFET trans-impedance amplifiers. 
A part of the energy of a pulse is deposited in the adjacent pixels, which contaminates another pulse. The optimum filtering is designed to mitigate this by reducing the weight for high-frequency content, in which the cross-talked pulse is more enhanced for its capacitive coupling nature.
The degradation is alleviated by applying the cross-talk cut at a sacrifice of the effective exposure time \cite{Mizumoto2022}. 

\begin{figure}[bt]
\centering
\includegraphics[width=0.8\linewidth]{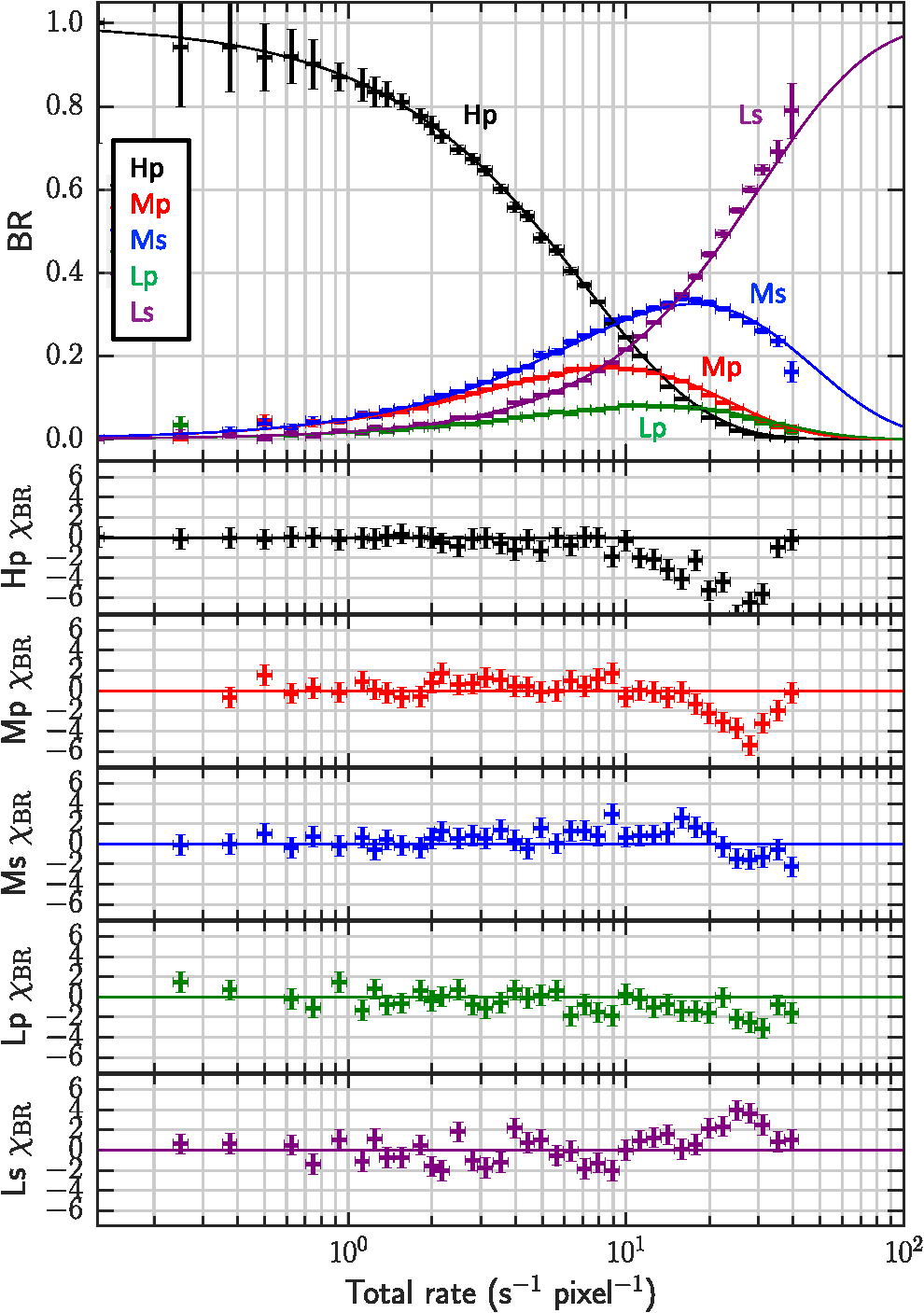}
\caption{%
The branching ratio (BR) is changed by event rates \cite{Ishisaki2018}.
In the top panel, the solid curves indicate the theoretical model by the Poisson statistics, and the data points are obtained with the SXS. 
 [Reproduced with permission from Ishisaki, Y., et al., Journal of Astronomical Telescopes, Instruments, and Systems, 4(1), 011217 (2018). Copyright 2018 Author(s), licensed under a Creative Commons Attribution 4.0 License.]}
\label{fig:branchingratio}
\end{figure}

The count rate influences the event grade. Figure \ref{fig:branchingratio} indicates a branching ratio of grades for a pixel. 
The branching ratios when the Crab nebula was observed with the SXS are shown, together with theoretical curves assuming the Poisson statistics for each grade. They are well matched \cite{Ishisaki2018}.
The higher the event rate is, the smaller the fraction of Hp is but the bigger the fraction of Ls is.

Note that if the secondary pulses are too close in time, they are not distinguished and are treated as one event, a so-called pile-up. The pile-up distorts the spectrum and the count rate. Since the pile-up events have a different trend from the normal ones, particularly in the pulse height vs. rise time relation, many of them can be removed \cite{Mizumoto2022}. 

The pulses are processed one by one from the stored data in the WFRB.
The {\it Resolve} has the requirement of processing up to a count rate of 200 s$^{-1}$ array$^{-1}$, including spurious events,  without event losses.
This is called the PSP limit. In case of a high count rate, the incoming photon rate can be beyond the PSP limit and events are discarded without processing \cite{Ishisaki2018}. The event lost is recorded as a pseudo-event and tagged in the event file. Therefore, the period of the event loss can be excluded.  

\subsection{Cooling chain}
\label{subsec:cooling}

\begin{figure}[b]
  \centerline{
    \includegraphics[width=0.9\textwidth]{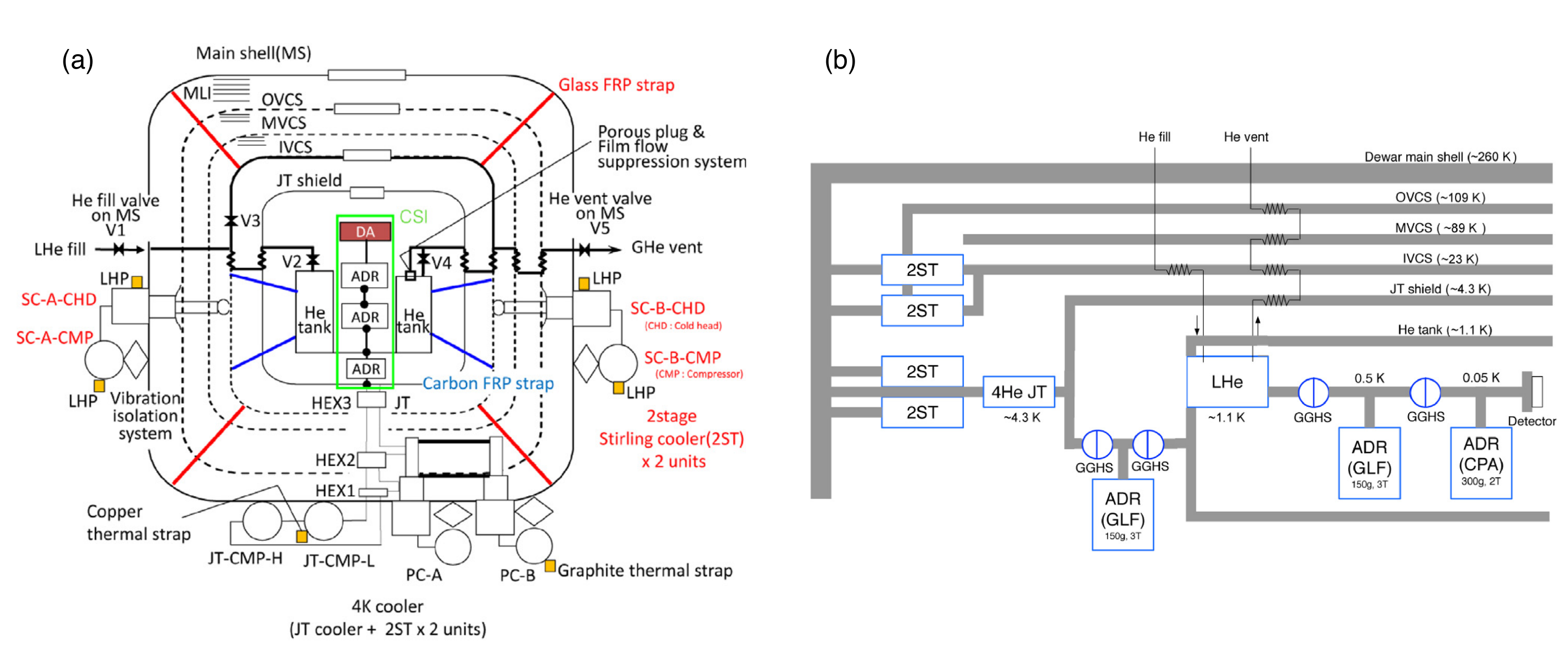}
    }
\caption{(a) Schematic view and (b) cooling chain of the {\it Resolve} cooling system \cite{Ezoe2020}. [Reproduced with permission from Ezoe, Y., Ishisaki, Y., Fujimoto, R., et al., Cryogenics, 108, 103016 (2020)]}
\label{fig:resolve-cooling}
\end{figure}

The X-ray micro-calorimeters of the SXS and the {\it Resolve} are operated at a low temperature of 50 mK to obtain the required energy resolution of $<7$ eV (FWHM) at 6 keV. Both the SXS and the {\it Resolve} use the cooling chain, schematically shown in Figure~\ref{fig:resolve-cooling}. 
The main objectives of the cooling system are
(1) to maintain the detectors at 50~mK with high duty cycle ($>90$\%), and (2) to satisfy the lifetime requirement of over 3 years. For (1), the cooling system is equipped with a 2-stage ADR and 30~L liquid helium (LHe) as the heat sink for the ADR. For (2), a ${}^4$He Joule-Thomson (JT) cooler unit and 2-stage Stirling (2ST) cooler units are adopted, which cool radiation shields to $\sim4.5$~K and $\sim20$~K, respectively, to reduce heat load to the He tank. As described in Section~\ref{sec:intro}, Suzaku XRS resulted in a failure of the instrument due to loss of LHe in orbit. Thus, the SXS and Resolve cooling systems were designed so that the ADR could continue to cool the detectors to 50~mK even after the LHe is depleted. This was accomplished by introducing another unit (stage-3) of ADR which could use the JT cryocooler as a heat sink.

After launch, the detector array and the calorimeter thermal sink are cooled by the 2-stage ADR to 50~mK \cite{Shirron2018}, which uses the LHe as a heat sink. These two stages are located in a well at the forward end of the He tank, and are arranged in series to make mechanical and thermal connections to the He tank. Whenever the magnet current of stage-1 ADR reaches zero, the ADR can no longer keep the detector temperature, and a recycle is needed. This takes about an hour, and will happen about every 44~hours in orbit. After the LHe is exhausted, the ADR changes operation to a cryogen-free mode  \cite{Sneiderman2018} in which the stage-3 ADR is continuously cycled to pump heat from the He tank to the JT cryocooler, while the stage-2 ADR is operated in a coordinated way to keep the He tank temperature stable at $\sim1.4$~K. Stage-1 recycles using stage-2 as its heat sink about every 16 hours in orbit; a process which also requires about an hour. Thus in both operating modes, the duty cycle for detector cooling to 50~mK is $>90$\%.

In order to achieve a LHe lifetime of at least 3 years, the heat load to the He tank must be kept below about 1~mW. This is accomplished by surrounding the tank with four vapor-cooled shields (VCS) called, in order from inside, JTS (Joule-Thomson shield), IVCS (Inner VCS), MVCS (Middle VCS), and OVCS (Outer VCS). The JTS is cooled by a $^4$He JT cryocooler \cite{Sato2021}, which uses two 2ST cryocoolers as precoolers (PC-A and PC-B). The OVCS and the IVCS are cooled by two 2ST cryocoolers \cite{Sato2014}, named shield cooler A and B (SC-A and SC-B). To keep LHe during ground tests and at launch, the components are contained in the dewar vacuum vessel, and its inside is maintained at low pressure on the ground. The compressors of the cryocoolers are mounted on the dewar main-shell (DMS).


\begin{figure}[b]
  \centerline{
    \includegraphics[width=0.9\textwidth]{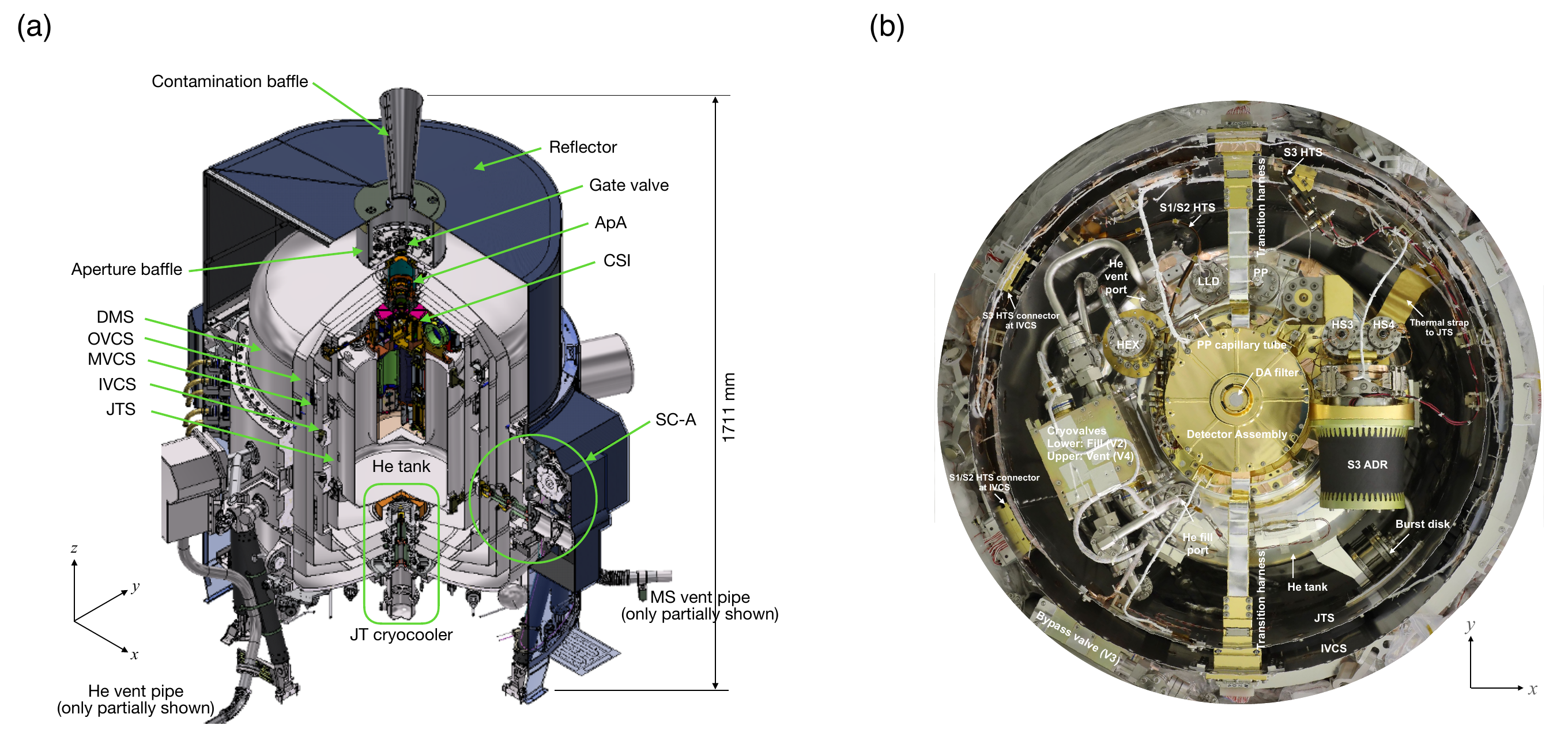}
    }
\caption{(a) A schematic view of the {\it Resolve} dewar and (b) a picture of the inside of the IVCS \cite{Ishisaki2022}. [Reproduced with permission from Ishisaki, Y., et al., Proc. SPIE Int. Soc. Opt. Eng., 12181, 121811S (2022)]}
\label{fig:resolve-dewar}
\end{figure}



\begin{figure}[tb]
  \centerline{
    \includegraphics[width=0.95\textwidth]{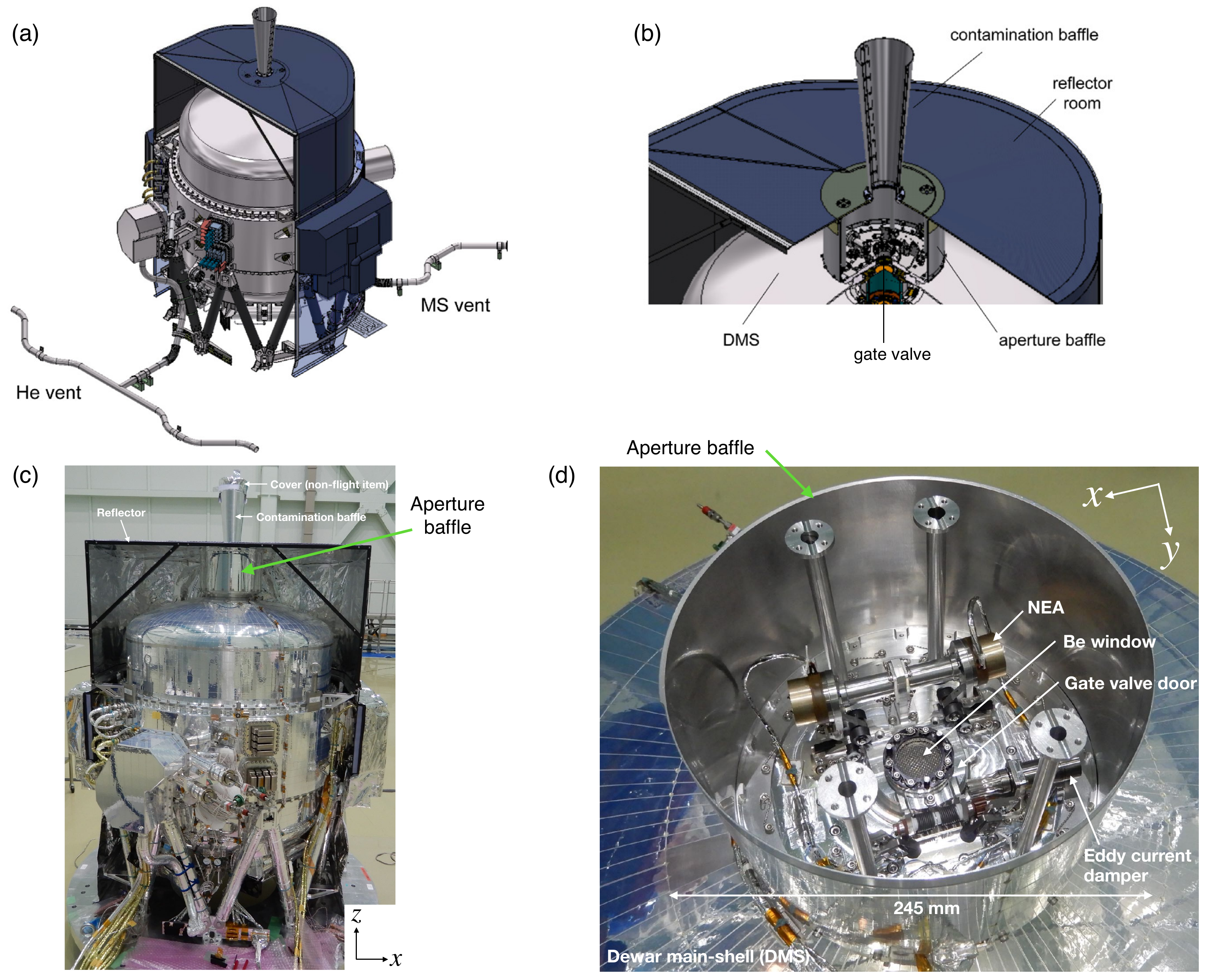}
    }
\caption{ Drawings of (a) the {\it Resolve} dewar and (b) the dewar topside \cite{Ezoe2020}. Pictures of (c) the {\it Resolve} flight dewar and (d) inside the aperture baffle \cite{Ishisaki2022}.[Reproduced with permissions from Ezoe, Y., Ishisaki, Y., Fujimoto, R., et al., Cryogenics, 108, 103016 (2020), and Ishisaki, Y., et al., Proc. SPIE Int. Soc. Opt. Eng., 12181, 121811S (2022)]}
\label{fig:dewar-img}
\end{figure}

When LHe is used in zero gravity, it is common to use a porous plug (PP) phase separator to retain the LHe in the tank while venting boil-off gas. SXS and Resolve are unusual in that the vent rate is about 30~$\mu$g\,s$^{-1}$, depending on the heat load on the He tank, which is the smallest flow rate among past space astronomy missions using LHe.
Since ${}^4$He will also leak out as a superfluid film around the perimeter of the PP, it is necessary to include a means to recapture as much of the latent heat of this film as possible, with a goal of reducing the film loss to less than 2~$\mu$g\,s$^{-1}$. A film flow suppression system, consisting of a capillary tube, a heat exchanger, and film flow killers, has been introduced \cite{Ishikawa2010}.
At the downstream side of the PP and the film flow suppression system, the helium gas is exhausted to outer space. An external vent pipe is attached to the dewar main-shell and routed to the outside of the spacecraft in order to prevent backflow of helium gas toward the spacecraft and dewar guard vacuum, which was the root cause of the failure of the XRS \cite{Kelley2007}. At the end of the vent line, the plumbing is divided into opposite directions to cancel the momentum of the vented gas, shown in Figure~\ref{fig:dewar-img} (a) \cite{Ezoe2020}. 



\subsection{Optical Chain}
\label{subsec:xray-path}

\begin{figure}[b]
    \centering
    \includegraphics[width=0.45\linewidth]{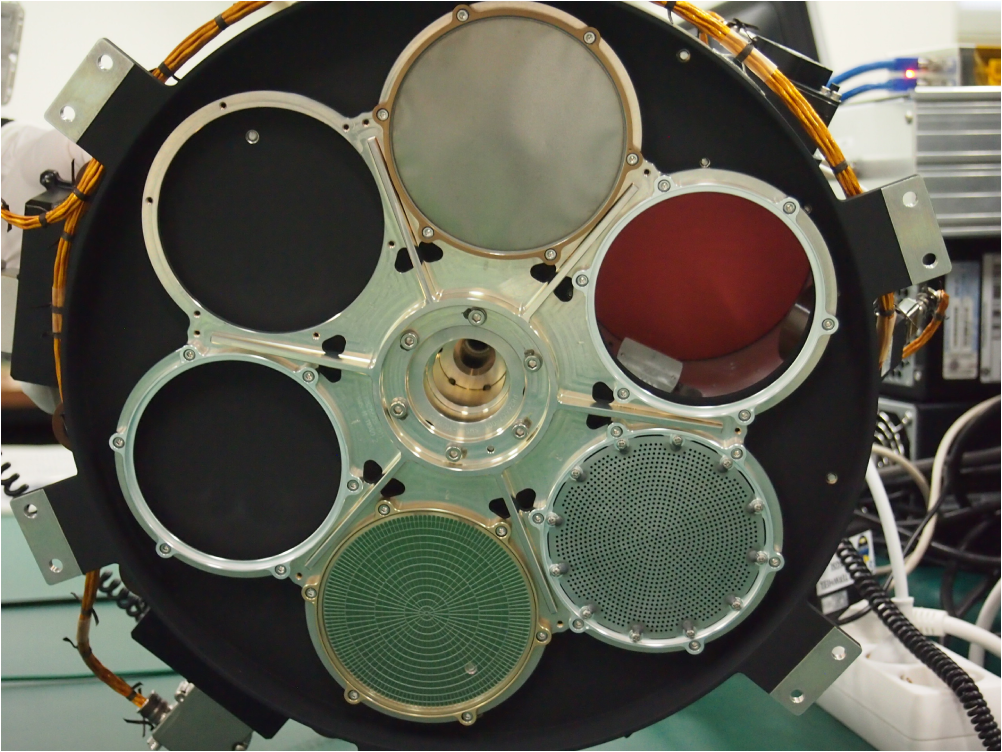}
    \includegraphics[width=0.45\linewidth]{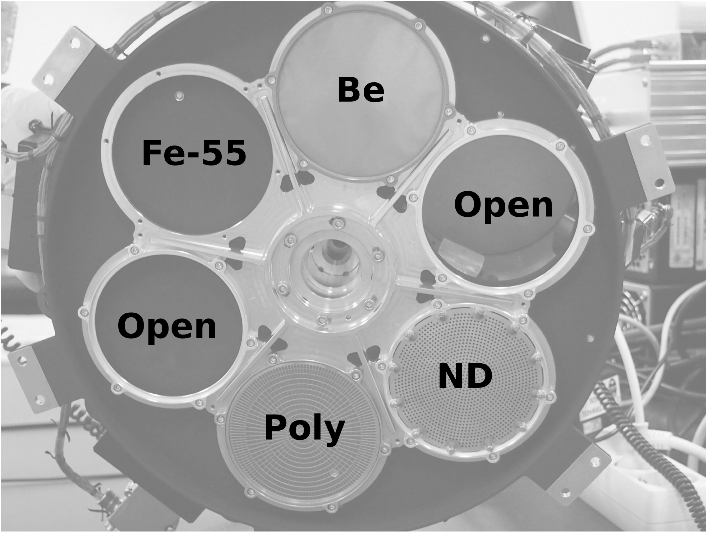}
    \caption{%
Pictures of the SXS FW \cite{deVries2018} [Reproduced with permission from de Vries, C. P., Haas, D., Yamasaki, N. Y., et al., Journal of Astronomical Telescopes, Instruments, and Systems, 4, 011204 (2018). Copyright 2018 Author(s), licensed under a Creative Commons Attribution 4.0 License.].
}%
    \label{fig:fw}
\end{figure}

X-rays collected by the mirror assembly go through some components. First, X-rays pass the FW to control the incident X-ray, then they reach the detector by passing through a GV, an ApA, and several filters.  

As shown in Figure \ref{fig:fw}, the FW has six positions. It rotates to set a filter position that a user selects.
The positions are assigned to ``Open'', ``Be'' filter, ``ND'' filter, ``Polyimide (OBF)'' filter, and ``$^{55}$Fe''.
The two positions of ``Open'', which mean blank, are our default position. ``Be'' filter consists of beryllium of a thickness of 27$~\mu\mathrm{m}$ and blocks the lower energy X-rays.
The neutral density (ND) filter is also blocking X-rays for bright objects. ``ND'' filter is made of a molybdenum sheet whose thickness is $0.25~\mathrm{mm}$ with a lot of $1.1~\mathrm{mm}$ diameter holes. The opening ratio is computed as about 24.5\%. The filter reduced the photons for the whole energy band.
``Polyimide'' filter is utilized to block contamination of the optical light.
For calibrations of the energy gain scale of the detector array, $^{55}\mathrm{Fe}$ sources are installed in the position ``Fe-55''.

The primary purpose of these filters is to reduce the photon count rate in an appropriate dynamic range of the intensity of bright sources. As described in Section \ref{subsec:characteristics}, there is a limitation for processing the signals so-called PSP limit. In addition, the incoming photon rate changes the event grade which is related to energy resolution.
Therefore, it is needed to suppress the number of incoming photons to the detector array. Guest observers of XRISM can choose two filters, ``ND'' and ``Be'' for blocking X-rays, in place of the ``Open'' position.

The GV is attached to the DMS at the X-ray aperture to seal the dewar guard vacuum on the ground.
The GV is planned to be kept closed for about one month after launch to avoid contamination on the DMS filter which is one of five thin filters installed in the X-ray incident path to prevent optical loading by photons larger than X-rays. For the ground test and observations in orbit prior to opening the GV, a beryllium window \cite{Midooka2021} with $\sim270$ $\mu$m thickness is installed in the GV door. 

\begin{figure}[tb]
\centering
\includegraphics[width=\linewidth]{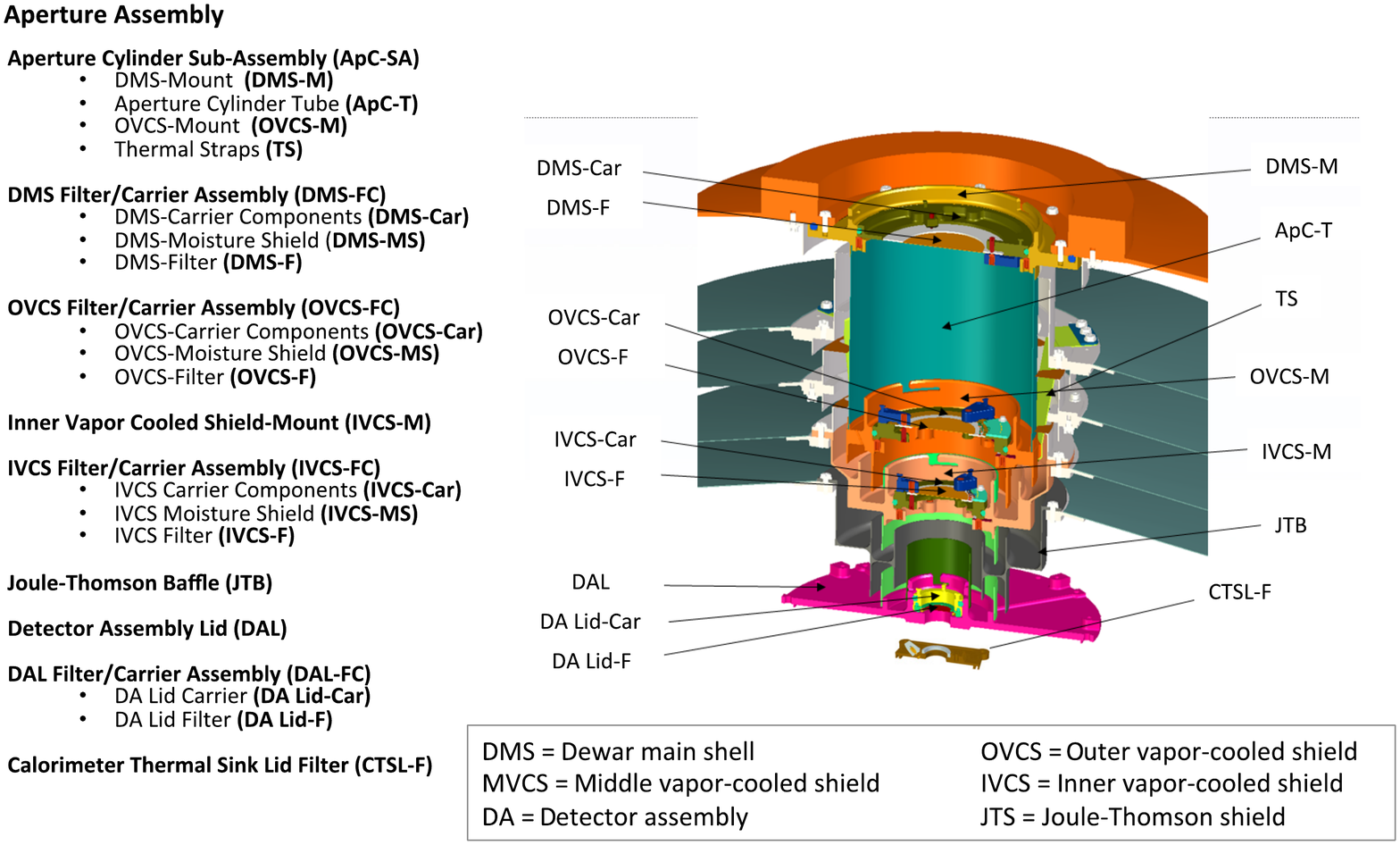}
\caption{Schematic view of the SXS aperture assembly and blocking filters \cite{Kilbourne2018c}. [Reproduced with permission from Kilbourne, C. A., Adams, J. S., Arsenovic, P., et al., Journal of Astronomical Telescopes, Instruments, and Systems, 4, 011215 (2018). Copyright 2018 Author(s), licensed under a Creative Commons Attribution 4.0 License.]}
\label{fig:aperture-fig}       
\end{figure}

The SXS and the {\it Resolve} need to operate under the stable environment of 50 mK in orbit in order to keep the high energy resolution \cite{Kilbourne2018c}.
Thus, the ApA protects the coldest stage from the thermal radiation in the higher-temperature stages in the dewar  and also prevents radiation from the outside.
The ApA provides five thin-film radiation-blocking filters for the thermal radiation from the equipment, and the optical/UV photons from the sky.
The ApA also includes the baffle and their support structures as shown in Figure \ref{fig:aperture-fig}. 

Since the area around the GV was open to space in the SXS design, there was a risk of damaging the DMS filter, the outermost filter, due to micro-meteoroid and orbital debris (MMOD) after opening the GV. If MMOD hits the DMS filter and makes a hole, radiation would penetrate inside and degrade the detector's performance. Therefore, a cylindrical aperture baffle made of aluminum is newly introduced around the GV area, which can reduce the probability of MMOD hit to the DMS filter to $<$0.03 per year at an altitude of 570 km \cite{Ishisaki2022}. The aperture baffle can also protect the DMS filter from the Earth albedo (optical light leak) and atomic oxygen other than MMOD. Figure \ref{fig:dewar-img} (b) (c) (d) show drawings of the {\it Resolve} dewar top and the aperture baffle, and pictures of the {\it Resolve} flight dewar and the inside of the aperture baffle, respectively \cite{Ezoe2020, Ishisaki2022}.

\subsection{Calibration sources}
\label{subsec:cal-sc}

In micro-calorimeter spectrometers, the calibration of the detector system is important because it is the characteristics of the detector response that provide the spectral information \cite{Eckart2018, Leutenegger2018, Porter2018}. Particularly, the energy gain scale and LSF are essential calibration items to represent the spectral feature of the incident X-ray photons for astrophysical investigations. Therefore, each pixel required its own gain scale calibration and the gain also needs to be characterized independently for each event grade. 

\begin{table}[t]
\caption{Calibration sources}
\label{tab:rad_source}
\begin{tabular}{llll}
\hline
Source Type & Irradiated area & Line & Operation \\ 
\noalign{\smallskip}\hline\noalign{\smallskip}
$^{55}\mathrm{Fe}$ source & Pixel 12 & fluorescent lines of Mn & always \\
``Direct'' sources of MXS & all pixels & fluorescent lines of Cu and Cr, & intermittent operation \\
 & &  and bremsstrahlung radiation & \\
``Indirect'' sources of MXS & all pixels & fluorescent lines of Al and Mg & intermittent operation \\
Fe-55 filter & all pixels & fluorescent lines of Mn & supplemental use \\
\hline\noalign{\smallskip}
\end{tabular}
\end{table}

The SXS and the {\it Resolve} detector system has three types of calibration sources as shown in Table \ref{tab:rad_source}. First, a calibration pixel is mounted as described in Section \ref{subsec:detector}, which is continually illuminated by a collimated $^{55}\mathrm{Fe}$ radioactive source. The calibration pixel allows for continuous monitoring of short-term gain drift and for correcting common drift among the pixels. Second, $^{55}\mathrm{Fe}$ source installed in the FW can illuminate the whole calorimeter array. Third, the SXS and the {\it Resolve} adopt a modulated X-ray source (MXS) \cite{deVries2018, Eckart2018}. MXS can generate characteristic X-rays using photoelectrons, which are created by UV lights from an LED and accelerated to targets with a high voltage, with an adjustable duty cycle \cite{Sawada2022}.

\begin{figure}[tb]
    \centering
    \includegraphics[width=0.5\linewidth]{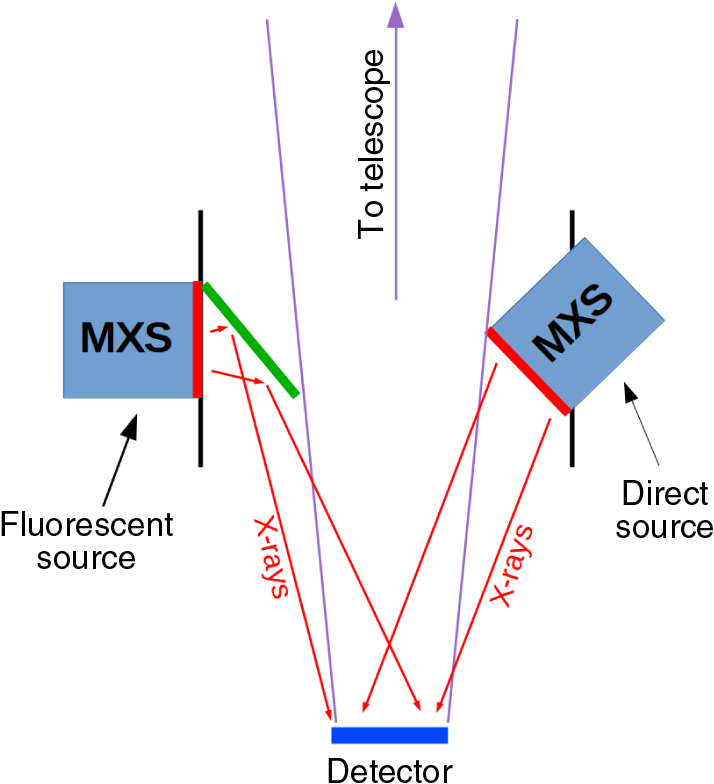}
    \caption{%
Schematic view of MXS \cite{deVries2018}. 
Characteristic X-ray lines of Cu and Cr and a continuum of the bremsstrahlung radiation are emitted from the direct source.
On the other hand, X-rays from the indirect source include Al and Mg fluorescent lines. [Reproduced with permission from de Vries, C. P., Haas, D., Yamasaki, N. Y., et al., Journal of Astronomical Telescopes, Instruments, and Systems, 4, 011204 (2018). Copyright 2018 Author(s), licensed under a Creative Commons Attribution 4.0 License.]
}%
\label{fig:mxsschematics}
\end{figure}

The MXS consists of two sources, called direct and indirect sources (see figure \ref{fig:mxsschematics}). 
Each source has a redundant system: nominal (NOM) and redundant (RED) sides. 
The X-rays from direct and indirect MXS exhibit different line spectra. The direct source generates Cu K$\alpha$ and Cr K$\alpha$ as well as the bremsstrahlung continuum, which provides the calibration references in the 5--8 keV band. The indirect source illuminates a secondary target to generate Mg K$\alpha$ and Al K$\alpha$ fluorescence line providing the references to trace occasionally low-energy gain scale in the 1--2 keV band \cite{Sawada2022}. The MXS irradiation can cover all pixels and pulsed X-ray emission from the MXS can be performed by switching the LED periodically. During observations of celestial targets, correction of the gain scale can be carried out using the MXS lines. X-ray events caused by the MXS can be separated from the X-rays of the celestial target using the time information of the MXS pulses. 

\section{Data processing and calibrations}
\label{sec:process}


\subsection{Offline processing of XRISM/{\it Resolve}}

XRISM data analysis starts by converting raw spacecraft packet telemetry received on the ground into the standard FITS format. This conversion of space packets to FITS format, along with the assignment of time, is performed in the pre-pipeline processing stage. This is followed by the pipeline processing stage where the data are calibrated and screened.

The pre-pipeline (PPL) represents the first step of the processing \cite{Terada2021}. Raw telemetry data are divided into First Fits Files (FFFs) that organize the data for distribution into the unfiltered event (uf). For the {\it Resolve}, the PPL (1) creates housekeeping (HK), calorimeter pixel event and anti-co event FFFs, (2) assigns time (with no barycentric correction), (3) divides files by filter wheel setting, and (4) constructs good time interval (GTI) files for excluding the lost event duration. The {\it Resolve} HK data such as temperatures, currents, and voltages are stored in the HK1 FITS file. The micro-calorimeter pixel event (but not the anti-co) data are divided into separate satellite slew and pointing files. Additional HK data including the detector templates, average pulses, and noise spectra at 1 k and 8 k resolutions are stored in the HK2 FITS file.

In the PPL, there are also common processes for all the instruments. One of these is the creation of the observation mode GTI. This divides the data into the time intervals before and after the attitude stabilization for each maneuver to an observation target, which are called slew and pointing durations, respectively. For the {\it Resolve}, the micro-calorimeter pixel event data are divided into slew and pointing in the PPL, while the anti-co event is not divided.


The pipeline (PL) processing applies event calibration and screening to FFFs, merging a number of GTI files for the latter purpose. Some of these, such as the GTI used to exclude times of ADR recycling are created in the PL. Users may fully reprocess an entire sequence of XRISM data using the HEASoft ftools, \texttt{xapipeline} or \texttt{rslpipeline}. The \texttt{rslpipeline} ftools applies all individual {\it Resolve} calibration and screening steps in the proper order and creates preview products. Reprocessing is recommended whenever a calibration or software update occurs, or when any non-standard processing is called for. The calibration steps include the assignment of coordinates (using the ftools \texttt{coordevt}) and energy (\texttt{rslpha2pi}), as well as the calculation of the gain history (see Section \ref{ssec:gain} below).


\subsection{Calibration files}
\label{ssec:caldb}

In order to derive accurate information from observation data of celestial objects of interest, the response of the instrument at the time of observation must be accounted for. The XRISM team (and the Hitomi team\footnote{Calibration files for Hitomi are available on the web page (\url{https://heasarc.gsfc.nasa.gov/docs/hitomi/calib/}). There are detailed description documents for each calibration file.}) provides this in the form of calibration files that are used directly by the data processing tools. 

\subsection{Pixel event data}
\label{ssec:eventdata}

After PL processing, the {\it Resolve} event FITS file columns for time, gain calibrated invariant pulse height, event grade, event status, lost event indicators, and so on are filled. Energy gain for each pixel is corrected based on gain histories created by the \texttt{rslgain} tool.
To avoid contamination of photons from celestial objects, it is important that flags recorded in an event file indicate coincidence with the anti-co detector, the MXS pulse timing, and cross-talk events. The column bits of the event \texttt{STATUS} column, populated by the \texttt{rslflagpix} tool, record these flags as indicated in Table \ref{tab:statusbit}.  \texttt{STATUS} bits are set by \texttt{rslflagpix} based on temporal coincidence within timescales defined in a calibration file (by default), or set by the user. Two types of electrical cross-talk are flagged based on the relative timing of events in other pixels with wiring proximity by applying relatively long and short intervals. 

\begin{table}[tb]
\begin{center}
\caption{List of \texttt{STATUS} bit$^{a}$}
\begin{tabular}{c|p{10cm}}\hline
\texttt{STATUS} bit & Description\\ \hline
0 & within general GTI \\
1 & within individual GTI for each PIXEL \\
2 & coincidence with anti-coincidence detector signal \\
3 & temporal proximity with another pixel \\
4 & coincidence with calibration pixel (12) within temporal proximity \\
5 & ``cross-talk''; coincidence with calibration pixel (12), and the energy is recoil energy \\
6 & ``cross-talk''; coincidence with adjacent wire pixel within short time scale \\
7 & ``cross-talk''; the pixel has maximum PHA in the events group within a short time scale \\
8 & ``MXS''; within MXS GTI (direct mode, including afterglow) \\
9 & ``MXS''; within MXS GTI (direct mode, considering only afterglow) \\
10 & ``MXS''; within MXS GTI (indirect mode, including afterglow) \\
11 & ``MXS''; within MXS GTI (indirect mode, considering only afterglow) \\
12 & ``cross-talk''; coincidence with adjacent wire pixel within long time scale \\
13 & ``cross-talk''; the pixel has maximum PHA in the events group within a long time scale \\
14 & reserved \\
15 & reserved \\ \hline
\end{tabular}
\label{tab:statusbit}
\end{center}
{\small $^{a}$ This table is written in fhelp of \texttt{rslflagpix}.}
\end{table}


\subsection{Time assignment}
\label{ssec:time}

The time of the micro-calorimeter events and anti-co events is first assigned onboard by the PSP based on the event trigger times. For X-ray grades of the micro-calorimeter pixel events (H, Mp, Ms, Lp, and Ls), the trigger times are further converted into photon-arrival times (calibrated event times) in the offline PPL process using the timing coefficients stored in a designated calibration file with the \texttt{rslsamcnt} and \texttt{xatime} tasks. There is no such offline event time conversion for the anti-co events, but the relative offset from the micro-calorimeter pixel events is calibrated and stored in the same calibration file. It is taken into account in the event flagging processes based on time-coincidence (see Section \ref{ssec:eventdata}). The purpose of the timing calibration can be divided into two: one is for the relative timing accuracy to perform the time-coincidence-based event flagging and screening and the other is for the absolute timing accuracy (requirement of 1 ms; Table \ref{tab:resolve-req}) desired from the scientific needs for time-domain astrophysical data analysis.

In the onboard time assignment, the trigger time for the H and M grades is set at which the time derivative of a pulse record exceeds a threshold, while for the L grade, it is set at which the pulse derivative reaches its maximum. This causes two major offsets in the trigger times: (1) a systematic delay for the L with respect to the H and M due to the different definitions of the trigger point and (2) an earlier trigger for higher-energy events for the H and M due to applying a uniform threshold for any X-ray energies. The latter is particularly large for low energy events below $\sim 1~\mathrm{keV}$ because the trigger point is closer to the maximum of the pulse derivative rather than its onset. This large energy dependence is mostly corrected in the onboard optimal filter processing. However, there is still a residual, smaller energy dependence caused by the non-linearity of the detector pulse. An actual detector pulse rises slightly faster for higher energy X-ray events, but the optimal filter uses a template evaluated at single energy ($5.9~\mathrm{keV}$ for the SXS and the {\it Resolve}). Thus, (2'-a) the onboard optimal filter overcorrects (undercorrects) the energy dependence for higher (lower) energy events, predicting slightly earlier (later) trigger times. For the L, although the time at the derivative maximum does not depend on much, (2'-b) there is also a small energy dependence on this trigger time. The relative offsets (1),  (2'-a), and (2'-b) require us to perform the relative timing calibration. In addition, (3) there is also a common delay to all the grades and energies originating from various factors such as a group delay in the analog detector pulse processing at the XBOX, which needs to be corrected as well to achieve the required absolute timing accuracy.

In the offline process, the relative and absolute timing offsets described above are corrected using the following equation:
\begin{align}
\mbox{}&[\mathrm{Calibrated\ Arrival\ Time}] \notag \\
\mbox{}& = [\mathrm{Trigger\ Time}] - \left(a\times\left(0.25\times\mathtt{RISE\_TIME}\right) + b\times \mathtt{DERIV\_MAX} + c \right), \nonumber
\end{align}
where \texttt{RISE\_TIME} is the time separation from the (original) trigger time to the time at which the pulse derivative reaches the maximum and \texttt{DERIV\_MAX} is the derivative value at the maximum. The coefficients $a$, $b$, and $c$ are defined separately for the different event grades and the micro-calorimeter pixels and stored in a designated calibration file. Both \texttt{RISE\_TIME}  and \texttt{DERIV\_MAX} can be used as a measurement of the X-ray energy. Thus, the coefficients a and b can be used to correct the residual energy dependence (2'-a) and (2'-b). Because of the overall pulse linearity (except a small non-linearity mentioned above), \texttt{RISE\_TIME}  is expected to change nearly proportional to \texttt{DERIV\_MAX}, and using either one of the coefficients a and b is sufficient to correct the residual energy dependence. For the SXS and the Resolve, only the coefficient b is used and the other one is kept at 0\@. The other relative offset, (1), is corrected by having different values of the c coefficient for the different grades. The absolute offset, (3), is corrected by having a common offset in c for all the grades and pixels. The above formalization of the timing conversion and coefficients also allows us to calibrate possible pixel-by-pixel offsets, although this type of offset should be negligibly small if changes in the trigger threshold between the template generation and actual observations are common to all the pixels.

\subsection{Gain correction and line spread function}
\label{ssec:gain}

\begin{figure}[b]
\centering
\includegraphics[width=0.6\linewidth]{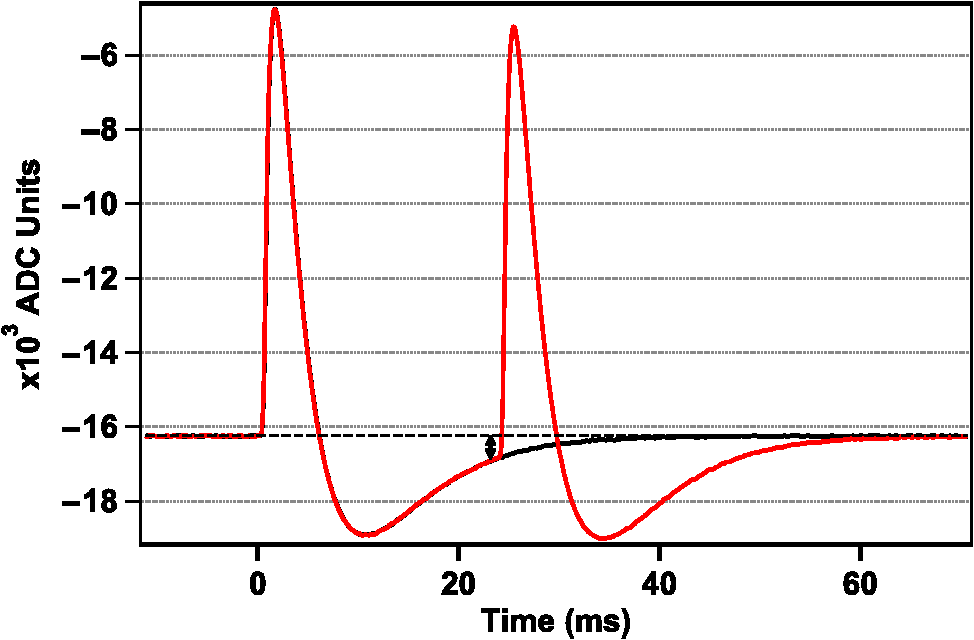}
\caption{Pulse profile of a Medium resolution event \cite{Eckart2018}. PSP gives the grades, Mp and Ms, to the first pulse and the second pulse, respectively. If the Ms grade pulse occurs before the Mp grade pulse could not be recovered completely to baseline, the evaluated pulse height of the Ms will underestimate. [Reproduced with permission from Eckart, M. E., Adams, J. S., Boyce, K. R., et al., Journal of Astronomical Telescopes, Instruments, and Systems, 4, 021406 (2018). Copyright 2018 Author(s), licensed under a Creative Commons Attribution 4.0 License.]
}
\label{fig:ms_pulse}
\end{figure}

The energy gain scale and LSF are calibrated on the ground and in orbit \cite{Eckart2018, Leutenegger2018}.
The calibration file for energy gain scale correction is prepared for three operation temperatures, $49~\mathrm{mK}$, $50~\mathrm{mK}$ and $51~\mathrm{mK}$, for each channel based on the calibration on the ground. Thus this indicates that the gain is different by the effective temperature of CSIs \cite{Porter2018}. In order to compensate for the energy drift during the observation due to the fluctuation of the temperature, it is necessary to correct the gain. The {\it Resolve} has three types of radiation sources for gain correction shown in Table \ref{tab:rad_source}.

The tool to estimate the gain drift is available as \texttt{rslgain} for the {\it Resolve} and \texttt{sxsgain} for the SXS. The output file is called the gain history file. By reading it as an option to the tool \texttt{rslpha2pi} for the {\it Resolve} (\texttt{sxspha2pi} for the SXS) to convert pulse height to energy, users can obtain event files with the energy corrected for temperature fluctuation.

As shown in Figure \ref{fig:ms_pulse}, there is the possibility to underestimate the pulse height for the Ms grade pulse with an optimum filter. The magnitude of the underestimation is different by the arrival time of the second pulse. The effect is that the energy resolution is getting worse for the Ms grade event. The pulse height can be recalculated by using the tool \texttt{rslseccor} (\texttt{sxsseccor}) with calibration files after reassigning the grade for the event with \texttt{rslsecid} (\texttt{sxssecid}).

\begin{figure}[b]
    \centering
    \includegraphics[width=0.8\linewidth]{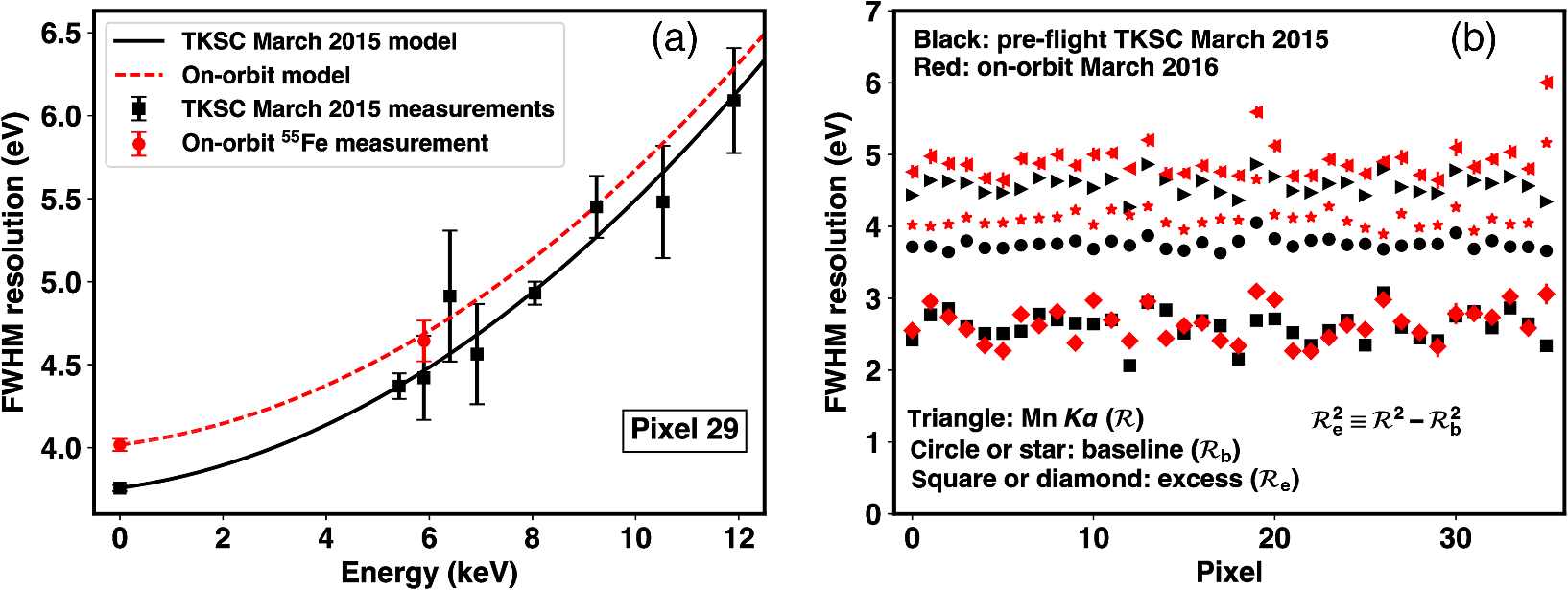}
    \caption{%
The energy resolution of the Gaussian core for SXS \cite{Leutenegger2018}.
Black points and lines show the data and model on the ground tests at Tsukuba Space Center (TKSC), respectively.
The red ones show the data and model in orbit.
These data were obtained by Hitomi/SXS. [Reproduced with permission from Leutenegger, M. A., Audard, M., Boyce, K. R., et al., Journal of Astronomical Telescopes, Instruments, and Systems, 4, 021407 (2018). Copyright 2018 Author(s), licensed under a Creative Commons Attribution 4.0 License.]
}%
    \label{fig:gausscore}
\end{figure}

The LSF is described with two components; Gaussian ``core'' and exponential ``tail'' as described in Section \ref{subsec:characteristics} \cite{Eckart2018, Leutenegger2018}. The Gaussian core is related to the detector noise, which depends on the operating environment whether on the ground or in orbit. For SXS, Figure \ref{fig:gausscore} shows the differences in the energy resolution of the core Gaussian between data taken on the ground and in orbit. The Gaussian core should be calibrated in orbit by using the calibration sources, the MXS and/or $^{55}\mathrm{Fe}$.
The energy resolution of the Gaussian core is modeled as 
\begin{equation}
\mathcal{R}(E) = \sqrt{\mathcal{R}_\mathrm{b} + \mathcal{R}_\mathrm{e}(E_\mathrm{ref})\left(\frac{E}{E_\mathrm{ref}}\right)}, \nonumber
\end{equation}
where $\mathcal{R}_\mathrm{b}$ is ``baseline'' resolution which is obtained by using the optimal filter for noise event and $\mathcal{R}_\mathrm{e}(E_\mathrm{ref})$ is the excess energy resolution at reference energy $E_\mathrm{ref}$.
The Gaussian LSF is evaluated for each pixel and contained in a calibration file, \texttt{rmfparam}. 
It would be changed over time.
On the other hand, the extended LSF is dependent on two different mechanisms and described by common parameters for all pixels. 
The mechanisms are escapes and electron loss continuum in the absorber (see Section 4.2.2 in \cite{Eckart2018}). 
The features are empirically modeled, which is contained in the calibration file. The tool \texttt{rslrmf} (\texttt{sxsrmf}) calculates the response matrix with considering LSF.

\section{Hitomi/SXS results}
\label{sec:hitomi-results}
\subsection{Hitomi/SXS performance in orbit}
\label{sec:sxs}

\begin{figure}[b]
  \centerline{
    \includegraphics[width=0.7\textwidth]{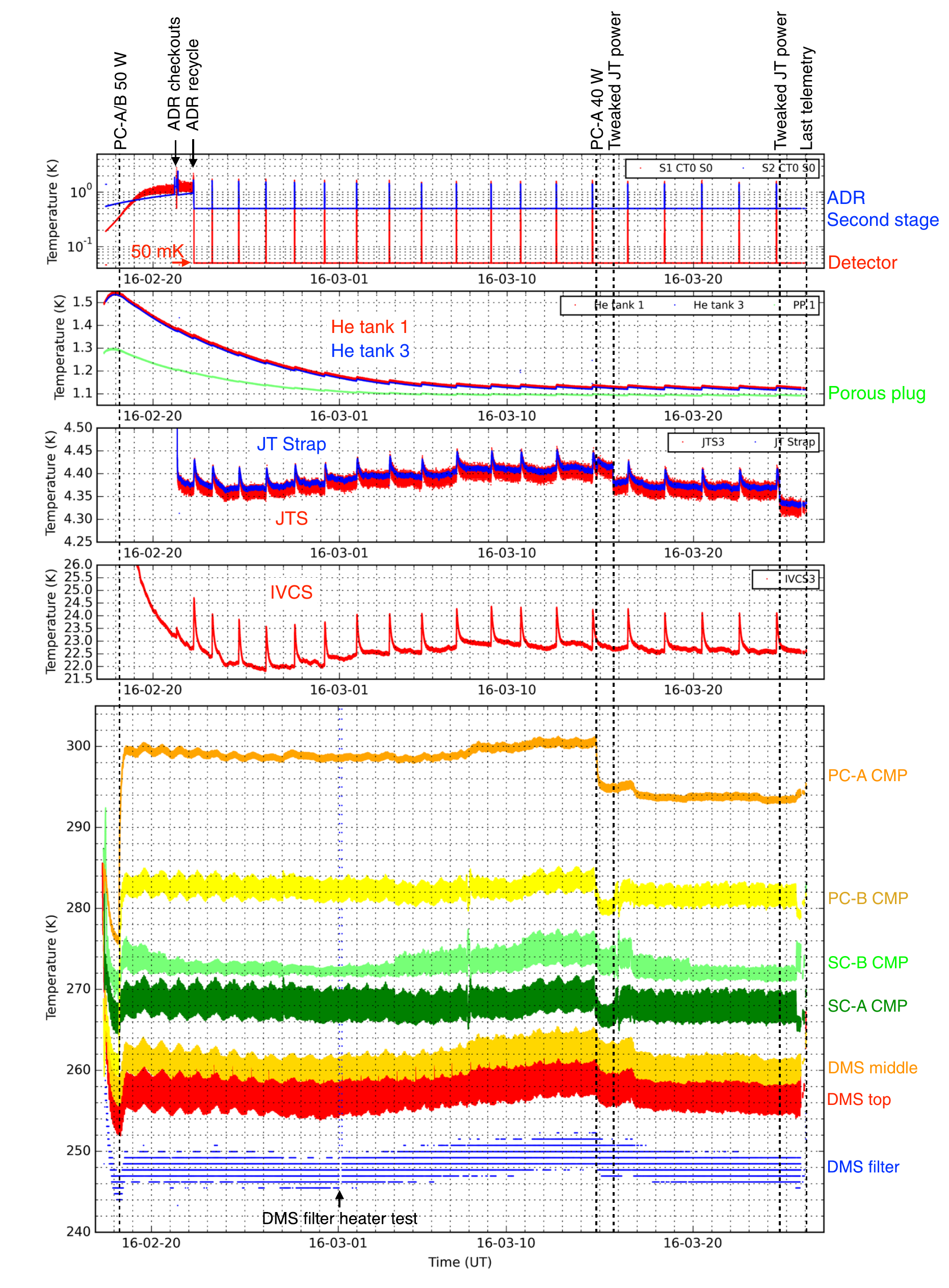}
    }
\caption{Temperature profiles of the SXS dewar in orbit \cite{Fujimoto2018}. [Reproduced with permission from Fujimoto, R., Takei, Y., Mitsuda, K., et al., Journal of Astronomical Telescopes, Instruments, and Systems, 4, 011208 (2018). Copyright 2018 Author(s), licensed under a Creative Commons Attribution 4.0 License.]}
\label{fig:sxs-temp}
\end{figure}

Hitomi was launched on February 17, 2016. During the rocket's acceleration, the He vent valve was opened to start venting helium gas through the PP 5 min after launch. After only about 20 minutes of launch, temperature monitoring of the He tank was started and normal venting was verified. And then, operations of SC and PC cryocoolers at low voltages were started, and an hour later the SC power was set at the nominal power (50 W$\times$2). On the next day (February 18), the PC was set at the nominal power, and the operation of the JT cryocooler started.
The JT power was increased gradually, and the JTS temperature reached 4.5 K on February 21. 
The ADR first recycle was carried out on February 21, and then the ADR recycles were performed periodically and the detector temperature was kept at 50 mK. Figure \ref{fig:sxs-temp} shows temperature profiles of the dewar until the last telemetry on March 26 \cite{Fujimoto2018}. The He tank temperature was slowly decreasing and reached $\sim$1.12 K at the end. A mass flow rate of the vented helium gas was estimated based on the ground test results to be about 34 $\mu$g s$^{-1}$, which corresponded to 0.70 mW heat load on the He tank \cite{Ezoe2017}. 
Temperature increases during ADR recycling can also be used to estimate the heat load on the He tank, and they were consistent \cite{Shirron2018}. 
Based on the estimated heat load, the lifetime requirement of 3 years would have been satisfied, even if degradation of the cryocoolers is considered.

The SXS was still in the commissioning phase when Hitomi lost attitude control. The in-orbit operation of the SXS, including the operations of the power system, cooling system, and signal acquisition system, was a complete success \cite{Tsujimoto2018}. For the cooling system to run detectors at 50 mK, the cooling chain worked successfully as mentioned above, and cooling is performed by the ADR and the temperature is kept within a few $\mu$ Kelvin by Proportional-Integral-Differential (PID) control in orbit \cite{Fujimoto2018}.
The vibration isolation system (VIS) for micro-vibration due to the mechanical cryocooler operations also worked as expected and the in-orbit detector spectral performance and cryocooler cooling performance were consistent with that on the ground \cite{Takei2018}. 
As a result, the SXS was confirmed to have comparable performance in the ground tests and in orbit, including the performance for data readout, and achieved a high energy resolution of 5 eV at 6 keV in orbit as shown in Figure \ref{fig:resolution} \cite{Porter2018}. On the other hand, because the SXS operations ended early in the commissioning phase, the on-orbit detector could not be fully calibrated. The GV was not opened and the performance below 2 keV could not be verified, nor could the MXS be operated. The measurement of the gain scale, which determines the absolute measurement of photon energy, and the measurement of the LSF were also insufficient because the entire SXS instrument had not in thermal equilibrium \cite{Leutenegger2018}. Thus, although the SXS was not fully verified for on-orbit performance, it achieved high energy resolution even in orbit and showed the potential for future missions using X-ray micro-calorimeters. The initial results of the SXS reaffirmed the importance of high energy resolution spectroscopic observations and emphasized the importance of the SXS recovery as described in Section \ref{subsec:hitomi-results}. Therefore, the recovery mission, XRISM has been proposed to achieve the key science goal of Hitomi as a joint project among JAXA, NASA, and ESA \cite{Tashiro2018}.

\subsection{Hitomi/SXS observations for the Perseus cluster}
\label{subsec:hitomi-results}

\begin{figure}[tb]
\centering
\includegraphics[width=0.5\linewidth]{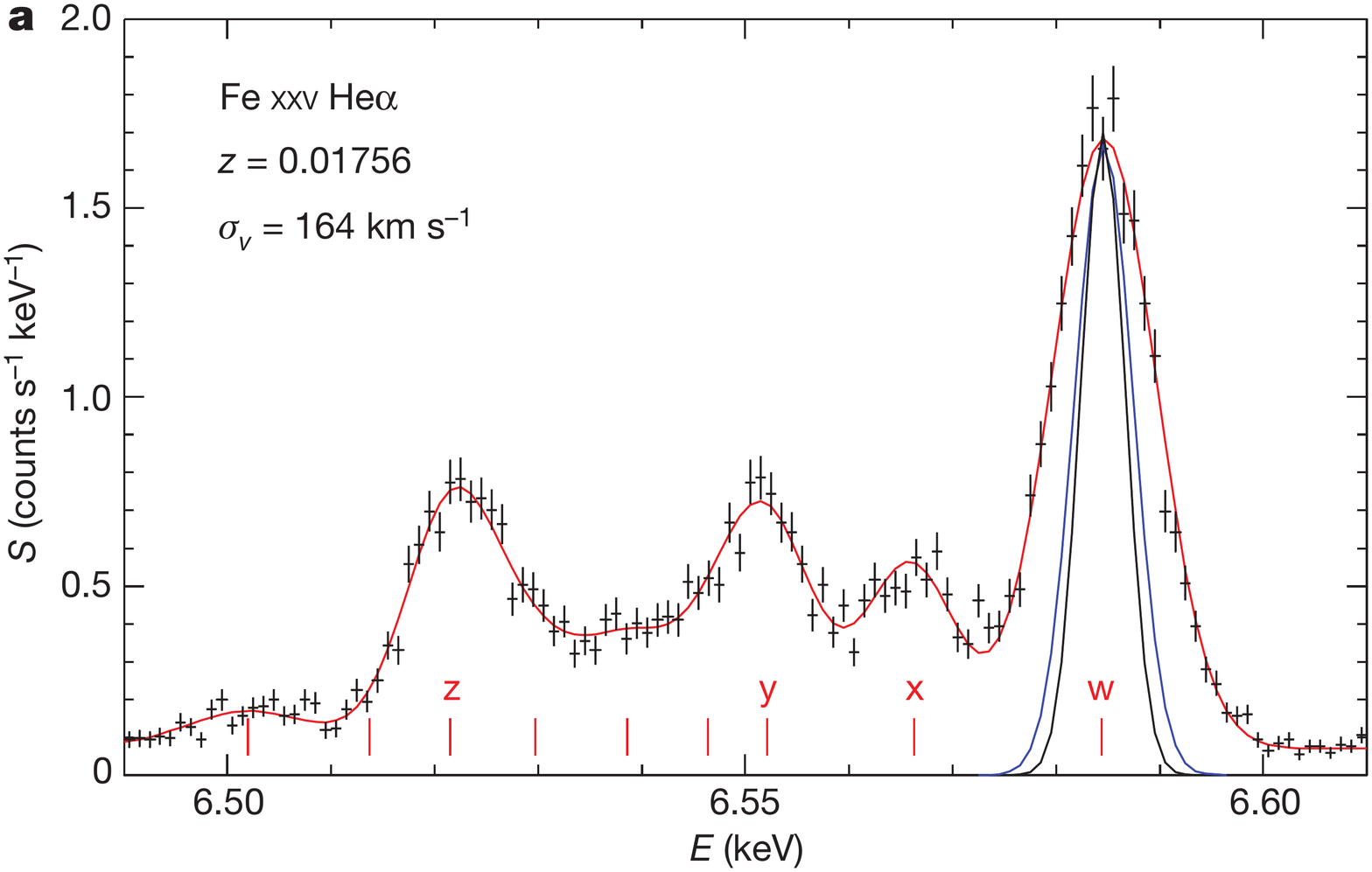}
\hfill
\includegraphics[width=0.48\linewidth]{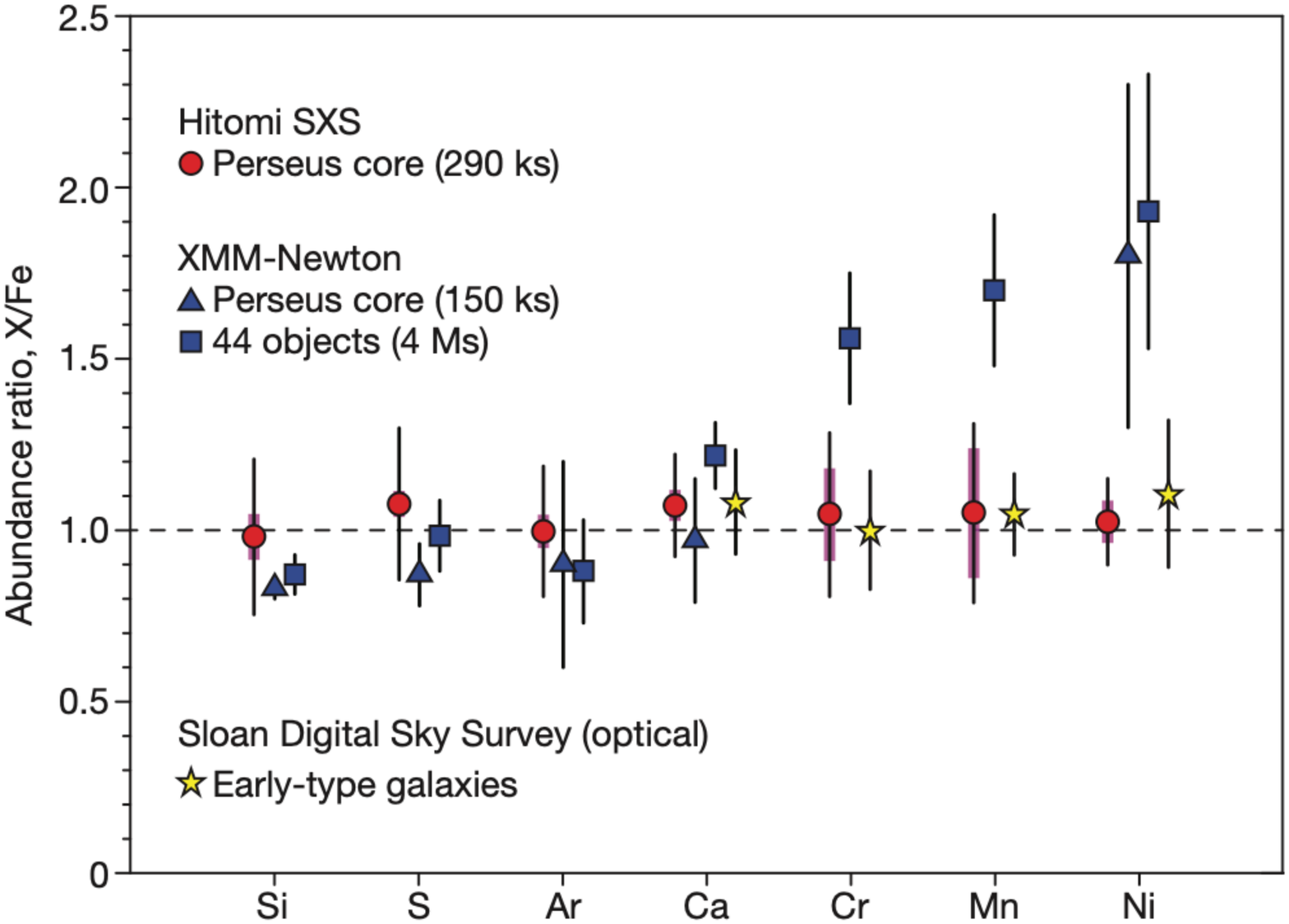}
\caption{(Left): SXS spectra of Fe$_{\rm XXV}$ He-$\alpha$ lines in the Perseus cluster core region \cite{Hitomi2016}. (Right): Metal abundance rations to Fe of the Perseus cluster observed by SXS \cite{Hitomi2017a}. [Reproduced with permissions from Aharonian, F., et al., [Hitomi], Nature 535, 117-121 (2016), and Aharonian, F., et al., [Hitomi], Nature 551, 478-480 (2017)]
}
\label{fig:hitomi-results}       
\end{figure}

Despite being in operation for only about a month, the SXS produced a very large number of important scientific results with high energy resolution. In particular, observations of the central region of the Perseus cluster of galaxies are reported in a series of papers \cite{Hitomi2016, Hitomi2017a, Hitomi2017b, Hitomi2018a, Hitomi2018b, Hitomi2018c, Hitomi2018d, Hitomi2018e}.
The SXS observation revealed fine structures such as Fe He-$\alpha$ emission lines for the first time. The observed velocity dispersion of $187 \pm 13$ and $164 \pm 10$ km s$^{-1}$ in the core and outer regions in the Perseus cluster, respectively, was much quieter than expected. Figure \ref{fig:hitomi-results} left shows the spectrum of Fe He-$\alpha$ in the core of the Perseus cluster, and the resonance line ($w$) is clearly wider than the instrumental broadening (blue line) and thermal broadening (black line). This result shows that the turbulent velocity in the Perseus cluster core does not affect the estimation of non-thermal effects on the mass estimates of galaxy clusters, which also affects the determination of cosmological parameters. The fine structure of the emission lines can now be clearly distinguished, and the ratio of resonant to forbidden lines in the iron emission lines due to resonant scattering has also been clarified, demonstrating for the first time that resonant scattering occurs at the center of the Perseus cluster. The SXS observation also gave the detection of not only the main strong lines from Si, S, Ar, Cr, and Fe but the weak lines from Cr, Mn, and Ni with high significance levels. These metal abundance ratios to Fe are similar to the Solar ratios as shown in Figure \ref{fig:hitomi-results}, and it constrained the metal enrichment mechanism of the Perseus cluster core. This result indicates that the elemental enrichment processes in our Galaxy and the Perseus Cluster are similar to each other, suggesting that the elemental cycles in the Universe would be universal. It is particularly important to note that observations of the Perseus cluster with the micro-calorimeter have achieved higher abundance determination accuracy than longer-duration observations with CCDs, despite the short observation time. The SXS demonstrated the power of high-energy spectroscopic observations and the importance of micro-calorimeter in high-energy astrophysics.

\section{XRISM/{\it Resolve} performance on ground test}
\label{sec:resolve}

\begin{figure}[b]
\centering
\includegraphics[width=\linewidth]{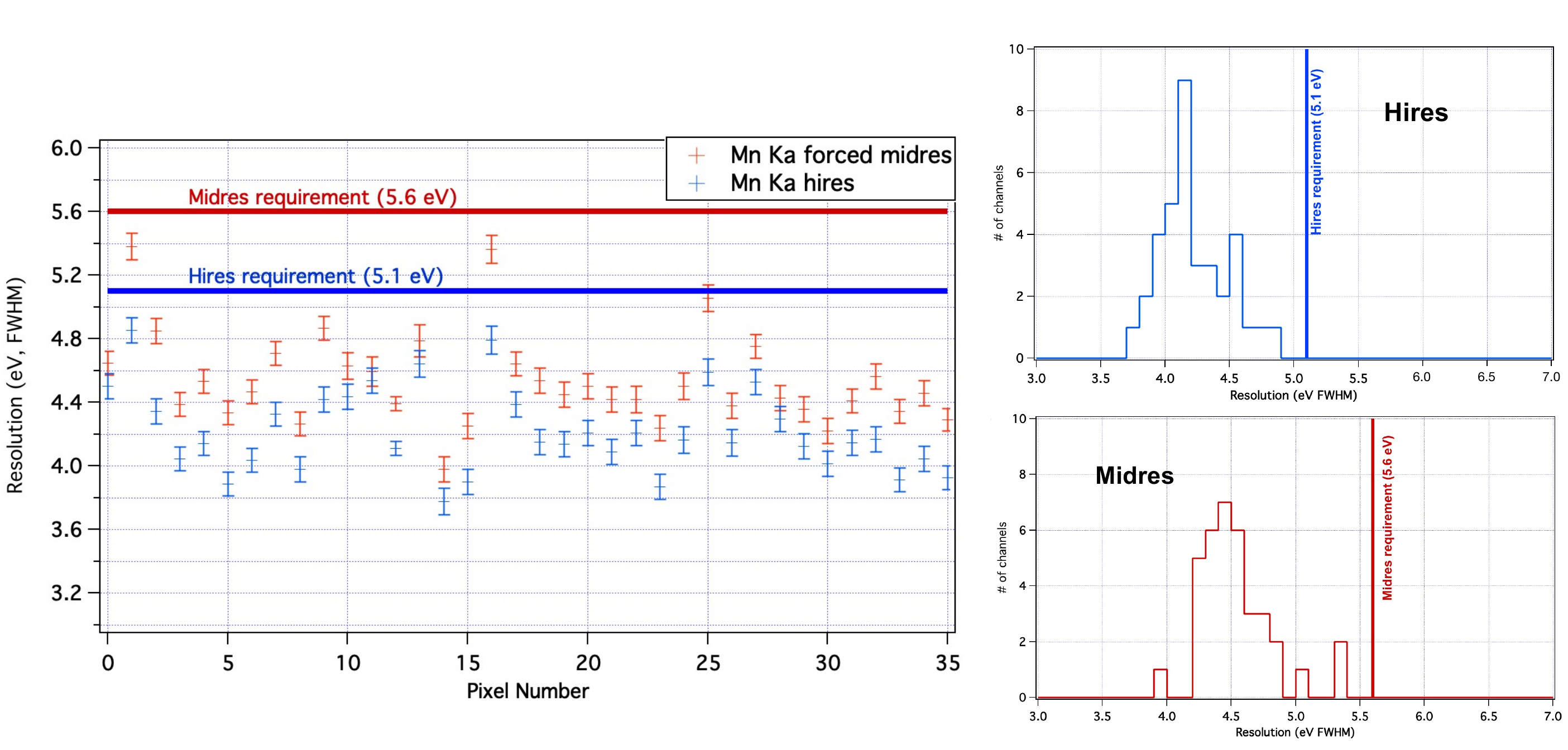}
\caption{(Left): Energy resolution (FWHM) of each pixel at 5.9 keV. (Right) Histograms of high and medium-resolution events for 36 pixels. \cite{Ishisaki2022}. [Reproduced with permission from Ishisaki, Y., et al., Proc. SPIE Int. Soc. Opt. Eng., 12181, 121811S (2022)]}
\label{fig:resolve-reso}       
\end{figure}

\begin{figure}[b]
\centering
\includegraphics[width=\linewidth]{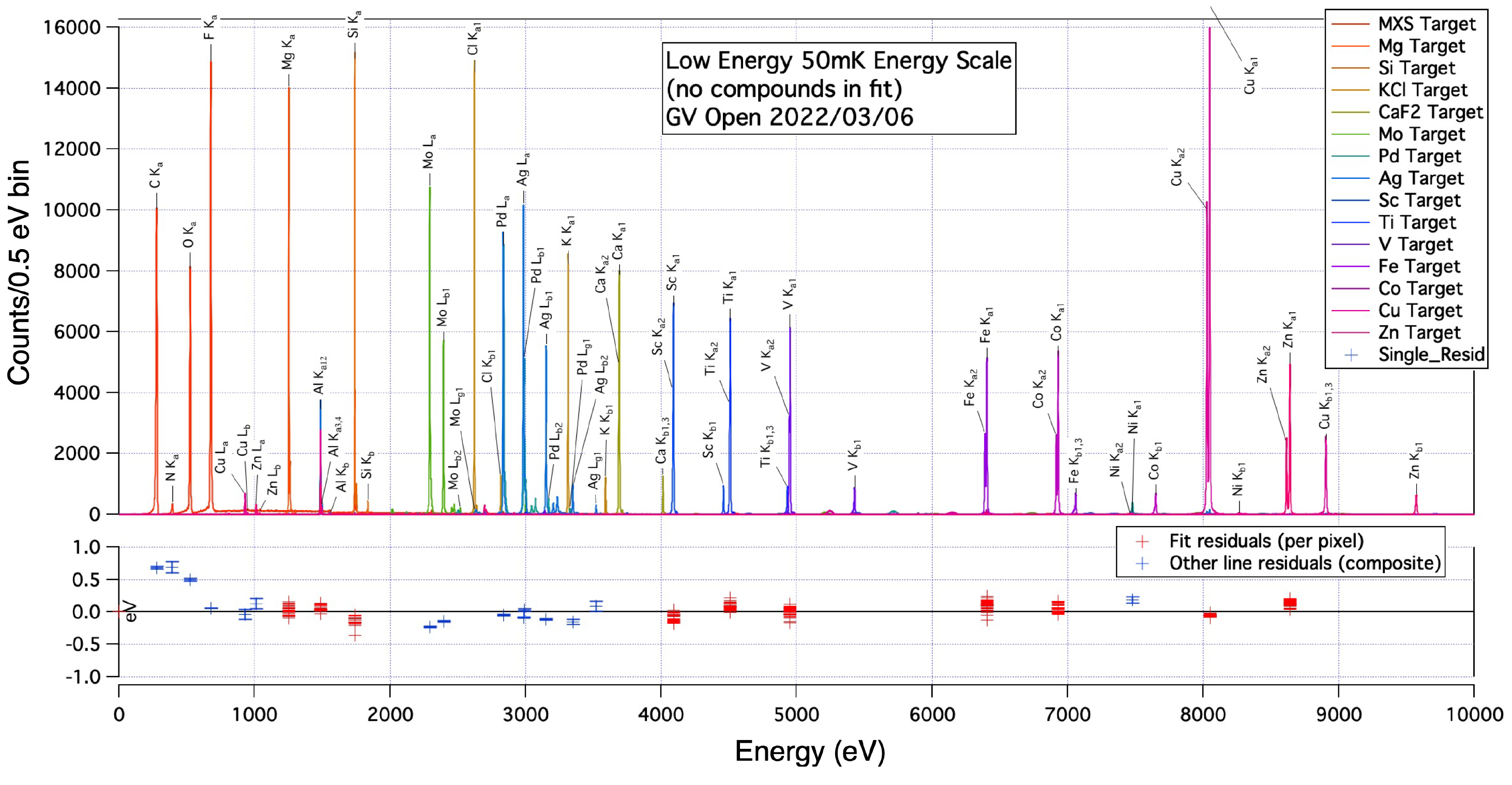}
\caption{Calibration of absolute energy scale accuracy for a number of characteristic X-rays with the gate valve open \cite{Ishisaki2022}. [Reproduced with permission from Ishisaki, Y., et al., Proc. SPIE Int. Soc. Opt. Eng., 12181, 121811S (2022)]}
\label{fig:resolve-gain}       
\end{figure}

Table \ref{tab:resolve-req} shows the requirements of the {\it Resolve} as same as the SXS. As for the {\it Resolve} performance, in the instrument level test in March 2022, the energy resolution for all the pixels met the requirement with a margin for high and medium resolution grade as shown in Figure \ref{fig:resolve-reso}. Figure \ref{fig:resolve-gain} shows a number of characteristic X-ray lines for the calibrations of the energy scale in which the pulse height is converted to the energy scale with a fit using a sixth-order polynomial function for each pixel. The absolute energy scale over 0.3 -- 9 keV energy range also met the requirement with margin. The calibration of the gain scale and line spread function in the 4 -- 25 keV energy range were also done both in cryogen mode and in cryogen-free mode, and those in the 0.3 -- 10 keV range in cryogen mode with the gate valve open. 

For the purpose of simulations of in-orbit conditions, the shield coolers were operated with a higher power to cool the IVCS, and the heat load of 0.55 mW on the He tank was obtained. Considering 10\% uncertainty and a conservative average heat load from the ADR, the heat load on the He tank at the beginning of life is expected to be 0.75 mW. Even if degradation of the cryocoolers (30\%) is included, the heat load is expected to satisfy the requirement of $<1$ mW at the end of life. Based on these results, the lifetime is roughly estimated to be 4.0 years for 35 L liquid helium. After launch, the heat load on the He tank can be estimated in the same way as the SXS, and the estimation of the life would be verified.

The ADR plays a key role in cooling the detector to 50 mK, and its performance directly affects the detector's performance. One of the critical functions of the ADR is to achieve high observing efficiency, which means that keeping 50 mK for an extended period of time in temperature-stable conditions is required. 
The requirement for temperature stability is 2.5 $\mu$K RMS (root mean square) or better during a 10 minutes period, in cryogen and cryogen-free modes. During the ground test, the operation algorithm and parameters were optimized. As a result, the performance of the temperature stability was 0.6 $\mu$K RMS on average in cryogen mode and 0.7 $\mu$K RMS in cryogen-free mode \cite{Ishisaki2022}.

In cryogen mode, the hold time which is the time to keep at 50 mK by one duty ADR cycle, was about 37.8 hours with the LHe temperature of typically about 1.24 K, then an operational duty cycle was $>$97\%. In cryogen-free mode, the hold time was about 16.7 hours. During the cryogen-free mode, the He tank was cooled to a stable 1.4 K using the stage-3 ADR. The operational duty cycle was obtained to be $>$93\% \cite{Ishisaki2022}. 
In the ground tests, the DMS temperature was higher than that expected in orbit as mentioned above. Therefore, the He tank temperature is expected to become lower to 1.12 K in orbit, and the smaller heat loads on the ADR will increase the hold time in cryogen mode. On the other hand, since in cryogen-free mode, the He tank temperature is controlled to keep the same at 1.4 K, the hold time is not affected. 

In order to keep tracking the gain drift in orbit, the MXS operation mounted on the FW is needed to illuminate the detector array in a pulsed mode at a duty cycle to be $\sim 1$\%\@. With the MXS operation, the MXS pulse-on time intervals are excluded in a GTI file, and it causes a small amount of decrease in observation efficiency. The MXS pulses also cause an increase in the instrumental (non-X-ray) background due to their exponential tails by each pulse and the loss of the throughput due to changes in the branching ratio of the event grade. Therefore, the calibrations for the pulse parameters on the ground has carried out to optimize the MXS operation in orbit, and the optimal solution for the MXS operation has been established \cite{Sawada2022}.

Also, the high count rate data sets are taken to evaluate the {\it Resolve} performance change \cite{Mizumoto2022}. There are three possible causes of spectral performance degradation: 1. CPU limit, 2. pile-up, and 3. electrical cross-talk. As for the CPU limit, the requirement of pulse processing is $>200$ s$^{-1}$ array$^{-1}$, including background and spurious events without event losses. 
The high count rate data set in the instrument level test on the ground showed that this requirement should be met.
The pile-up distorts the spectrum and count rate due to the arrival times of the primary and secondary pulse events being too close (see also Section \ref{subsec:event_processing}). Electrical cross-talk is another possible cause of energy resolution degradation. When one pulse arrives, a part of its energy is deposited into adjacent pixels, which in turn affects another pulse that would normally be unrelated. The behavior for the high count rate event is well studied and modeled with the data set in the instrument level tests on the ground \cite{Mizumoto2022}.

After the instrumental level test, the {\it Resolve} was integrated into the XRISM satellite in April 2022. The XRISM has conducted initial electrical and thermal vacuum tests and has confirmed that there is no change in the {\it Resolve} performance. The data sets for calibrations are also taken to develop a plan for MXS operation in orbit and to estimate the influence of high count rate observations. The XRISM will be transported to the Tanegashima Space Center for launch after mechanical environmental tests and final electrical tests.

\begin{acknowledgement}
This work was made possible by the collaborative efforts of all members of the Resolve team, SHI, and NEC engineers, which we greatly appreciate. Particularly, the authors are grateful to M. Eckart, C. Kilbourne, R. Kelley, M. Lowenstein, R. Shipman, P. Shirron, Y. Ishisaki, R. Fujimoto, M. Tsujimoto, M. Sawada, and Y. Takei for valuable comments and suggestions to our manuscript.
\end{acknowledgement}

\end{document}